\documentclass[12pt,draftclsnofoot,onecolumn]{IEEEtran}

\usepackage{amsmath,amsfonts}
\usepackage{algorithmic}
\usepackage{array}
\usepackage[caption=false,font=normalsize,labelfont=sf,textfont=sf]{subfig}
\usepackage{textcomp}
\usepackage{stfloats}
\usepackage{url}
\usepackage{verbatim}
\usepackage{graphicx}
\usepackage{cite}
\usepackage{graphicx}
\usepackage{epstopdf}
\usepackage{amssymb}
\usepackage[ruled, linesnumbered]{algorithm2e}

\usepackage{multirow} 

\usepackage{amsmath}

\usepackage{xcolor}

\newtheorem{myDef}{Definition}
\newtheorem{theorem}{Theorem}
\UseRawInputEncoding
\begin{document}

\title{Design of Two-Level Incentive Mechanisms for Hierarchical Federated Learning}

\author{Shunfeng~Chu,
        Jun~Li, 
        Jianxin~Wang,
        Kang~Wei, 
        Yuwen~Qian,
        Kunlun~Wang, 
        Feng~Shu, 
        and~Wen~Chen

\thanks{S. Chu, J. Li, J. Wang, Y. Qian, and F. Shu are with the School of Electronic and Optical
Engineering, Nanjing University of Science and Technology, Nanjing 210094, China. F. Shu is also with the School of Information and Communication Engineering, Hainan University, Haikou, 570228, China (e-mail: shunfeng.chu@njust.edu.cn; jun.li@njust.edu.cn; wangjxin@njust.edu.cn; admon@njust.edu.cn; shufeng0101@163.com).}
\thanks{K. Wei was with the School of Electrical and Optical Engineering, Nanjing University of Science and Technology, Nanjing 210094, China. He is now with the Department of Computing, The Hong Kong Polytechnic University, Hong Kong 999077, China. (e-mail: kangwei@polyu.edu.hk).}
\thanks{K. Wang is with the School of Communication and Electronic Engineering, East China Normal University, Shanghai 200241, China (e-mail: klwang@cee.ecnu.edu.cn).}
\thanks{W. Chen is with the Department of Electronic Engineering, Shanghai Jiao Tong University, Shanghai 200240, China (e-mail: wenchen@sjtu.edu.cn).}
}


\markboth{Draft}%
{Shell \MakeLowercase{\textit{et al.}}: Bare Demo of IEEEtran.cls for IEEE Journals}

\maketitle

\begin{abstract}
Hierarchical Federated Learning (HFL) is a distributed machine learning paradigm tailored for multi-tiered computation architectures, which supports massive access of devices' models simultaneously.
To enable efficient HFL, it is crucial to design suitable incentive mechanisms to ensure that devices actively participate in local training. In this paper, we design two-level incentive mechanisms based on game theory for HFL in a device-edge-cloud coordinating architecture, aiming at encouraging the participation of entities in each level. In the design of lower-level incentive mechanism, we propose a coalition formation game to optimize the device-edge association and bandwidth allocation. To be specific, we first develop efficient coalition partitions based on preference rules for optimizing device-edge association, which can be proven to be stable by constructing a potential function. Then, we develop a gradient projection method to optimally allocate bandwidth among the coalitions. In the upper-level game, we design a Stackelberg game algorithm to jointly maximize the utilities of the cloud and each edge server. The proposed algorithm is able to determine the optimal number of aggregations at each edge server, as well as the reward provided by the cloud server to the each edge server for the performance improvement due to the edge aggregations. Numerical results indicate that the proposed two-level incentive mechanisms can achieve better performance than the benchmark schemes. Moreover, our proposed methodology has been shown to achieve accuracy improvements of up to 3\% on real datasets, e.g., Cifar-10, in comparison to other benchmarks.
\end{abstract}

\begin{IEEEkeywords}
Hierarchical federated learning, Game theory, Coalition game, Resource allocation, Stackelberg game
\end{IEEEkeywords}

\section{Introduction}
\IEEEPARstart{R}{ECENTLY}, Artificial Intelligence (AI) technology is rapidly advancing and finding increasing applications in various fields \cite{lecun2015deep,wei2020federated1}. The technology allows machines to possess cognitive abilities that are similar to humans, enabling them to process large volumes of data, make independent decisions, and adapt to their environments. For instance, machine translation and speech recognition technologies have become increasingly popular in the area of natural language processing, and are capable of achieving relatively accurate results. Voice assistants have become indispensable helpers in people's daily lives. In the field of computer vision, AI is capable of recognizing faces and actions, providing strong support for image recognition, autonomous driving, and other related domains.

Typically, traditional machine learning requires collecting a large amount of training data on a central server for model training \cite{rastogi2018privacy}. However, this approach not only incurs significant communication overhead but also carries the risk of user privacy leaks. Nowadays, the traditional central machine learning has been subject to increasingly strict policies such as the General Data Protection Regulation (GDPR) \cite{lim2020federated}. The deployment of data-driven AI is significantly impeded by the increasing restrictions on sharing training data with external parties \cite{yuan2023amplitude}. To address this critical issue, Google has proposed a new type of distributed machine learning called Federated Learning (FL), which trains AI models in a privacy-preserving manner \cite{wei2022user,li2022blockchain}. The core idea of FL is to update AI models by transmitting model parameters rather than local data samples of devices. Hence, FL enables efficient preservation of data privacy for device users since only the local update (e.g., parameter gradients) is sent to the central server via encrypted communication, instead of the local data \cite{deng2022blockchain}.

In the future, FL networks are envisioned to comprise thousands of heterogeneous IoT and mobile devices, which will pose some challenges to traditional FL techniques. The first issue is that the enormous communication overhead will become a critical bottleneck for FL \cite{lim2020hierarchical}. In traditional FL, there is a significant communication overhead as each participant needs to frequently transmit model or gradient parameters to the central server. Various solutions have been proposed, including model compression techniques like quantization and subsampling, and client selection \cite{wei2022low}, to address these issues. Nevertheless, despite the above measures, communication inefficiencies can still cause device dropouts and stragglers during the FL process\cite{yang2021energy}. Besides, devices at geographically distant locations may not be able to participate in FL training, which can adversely affect the generalization performance of model \cite{kang2020reliable}.

In order to address above issues, a new framework named hierarchical FL (HFL) has been proposed, where devices only need to upload their local model parameters to edge servers, such as small base-stations (SBSs) in cellular networks or access points in wireless networks \cite{luo2020hfel}. Each edge server can obtain the intermediate parameters by aggregating local models from devices it associated with. Then, the intermediate parameters are uploaded to the central server, such as the cloud server for further global aggregation. The advantages of the HFL architecture are fourfold. First, HFL can greatly reduce the number of costly global communication rounds with the remote central server. Then, the edge servers as relays can also reduces the dropout rate of devices whose communications resources are starved. In synchronous HFL, some scattered devices could not affect the global training performance of HFL. Next, the intermediate aggregation in the HFL framework also improves the personalized learning performance of FL. Finally, the convergence of HFL has been proven and empirical studies have shown that there is no significant reduction in model accuracy compared to conventional FL implementation \cite{abad2020hierarchical}.

However, there is still a significant learning latency in wireless HFL due to traffic load imbalance and limited wireless resources. To improve the efficiency of HFL systems in wireless networks, it is necessary to schedule the devices involved in training and optimize wireless resource. The work in \cite{mhaisen2021optimal} proposed solutions to improve the learning performance of HFL with non-IID training data by optimizing the user-edge assignment, analyzed the effect of data distribution on the learning performance and designed an optimal user-edge assignment strategy. The work in \cite{luo2020hfel} proposed a formulation for the joint allocation of computation and communication resources, as well as client association, in order to minimize energy consumption and delay during each communication round of model training in the HFL network. In \cite{zhao2022drl}, a deep reinforcement learning based joint resource allocation method has been proposed to achieve a more accurate model and reduce overhead for MEC-assisted HFL in IIoT. The work in \cite{he2022resource} proposed a DDPG-based solution that addresses the problem of dynamic association between devices and stations as well as the resource allocation in order to minimize energy consumption within a limited delay in the HFL system. The work in \cite{wen2022joint} formulated an optimization problem to design a joint helper scheduling and wireless resource allocation scheme, the problem simultaneously captured the uncertainty of wireless channels and the importance of the weighted parameters of HFL.

Nevertheless, the aforementioned works have ideally assumed that all devices are willing to participate in the training of HFL. Since local model training will incur significant costs for the involved devices in HFL, selfish devices may have no motivation to participate in HFL without proper incentive mechanisms. Currently, there is a scarcity of research on the design of incentive mechanisms for HFL, with most existing works focusing on incentive mechanism design for conventional FL. The work in \cite{huang2022collaboration} proposed a novel analytic framework for incentivizing effective and efficient collaborations for participant-centric FL, and designed efficient algorithms to achieve equilibrium solutions. The work in \cite{Shyuan2022reputation} proposed the reputation-aware hedonic coalition formation game to improve the sustainable efficiency of the FL system while taking into account the incentive design for devices' marginal contributions in FL system. The work in \cite{kang2019incentive} proposed an incentive mechanism based on contract theory to encourage FL workers to participate in global model training. The work in \cite{Lim2022decentralized,dynamic2021} considered designing incentive mechanisms in the HFL framework, and then proposed game-theoretic frameworks for the lower-level devices and upper-level servers, respectively. However, how to optimize the training performance and efficiency of a realistic HFL framework with constrained computation and communication resources remains a largely unexplored area.

In this paper, we incorporate a device-edge-cloud system based on HFL into commercial scenarios to encourage the participation of entities in each level. Two-level incentive mechanisms based on game theory are proposed for HFL in a device-edge-cloud coordinating architecture. In the lower tier of the HFL network, we formulate the device-edge association and bandwidth allocation problem as a coalition game, referred to as the lower-level game. Additionally, we employ Stackelberg game to devise a trading mechanism between the cloud servers and edge servers in the upper-tier network of HFL, referenced to as the upper-level game.
The main contributions of this paper are as follows:
\begin{enumerate}
  \item We incorporate the HFL network with multiple devices and edge servers into commercial scenarios and design two-level incentive mechanisms with a two-tiered computing structure to encourage the participation of entities in each tier in HFL.
  \item In the lower-level game, we formulate the joint device-edge association and bandwidth allocation problem as a coalition formation game. The coalition formation game is proved to be an exact potential game for coalition altruistic preference rule, which has at least one stable coalition partition. Besides, we propose the gradient projection method to solve the bandwidth allocation among the coalitions.
  \item In the upper-level game, we design a Stackelberg game algorithm to maximize the utilities of the cloud and each edge server. The proposed algorithm is able to determine the optimal number of aggregations at each edge server, as well as the rewards provided by the cloud server to the edge servers for the performance improvement due to the edge aggregations.
  \item Furthermore, numerical results indicate that the proposed algorithms can achieve better performance than the benchmark schemes. Our proposed methodology has been shown to achieve accuracy improvements of up to 3\% on real-world datasets in comparison to other benchmark schemes.
\end{enumerate}

The rest of this paper is organized as follows. Section \ref{sec:System Model} describes the system model. Coalition formation game is formulated
in Section \ref{sec:Coalition formation game formulation}. Section \ref{sec:Upper-Level Stackelberg Game} describes the Upper-Level Stackelberg Game.
Section \ref{Simulation Results and Discussion} presents the simulation results. Finally, Section \ref{sec:conclusion} concludes this paper.


\begin{figure}
\center
\includegraphics[height=6.5cm,width=8.7cm]{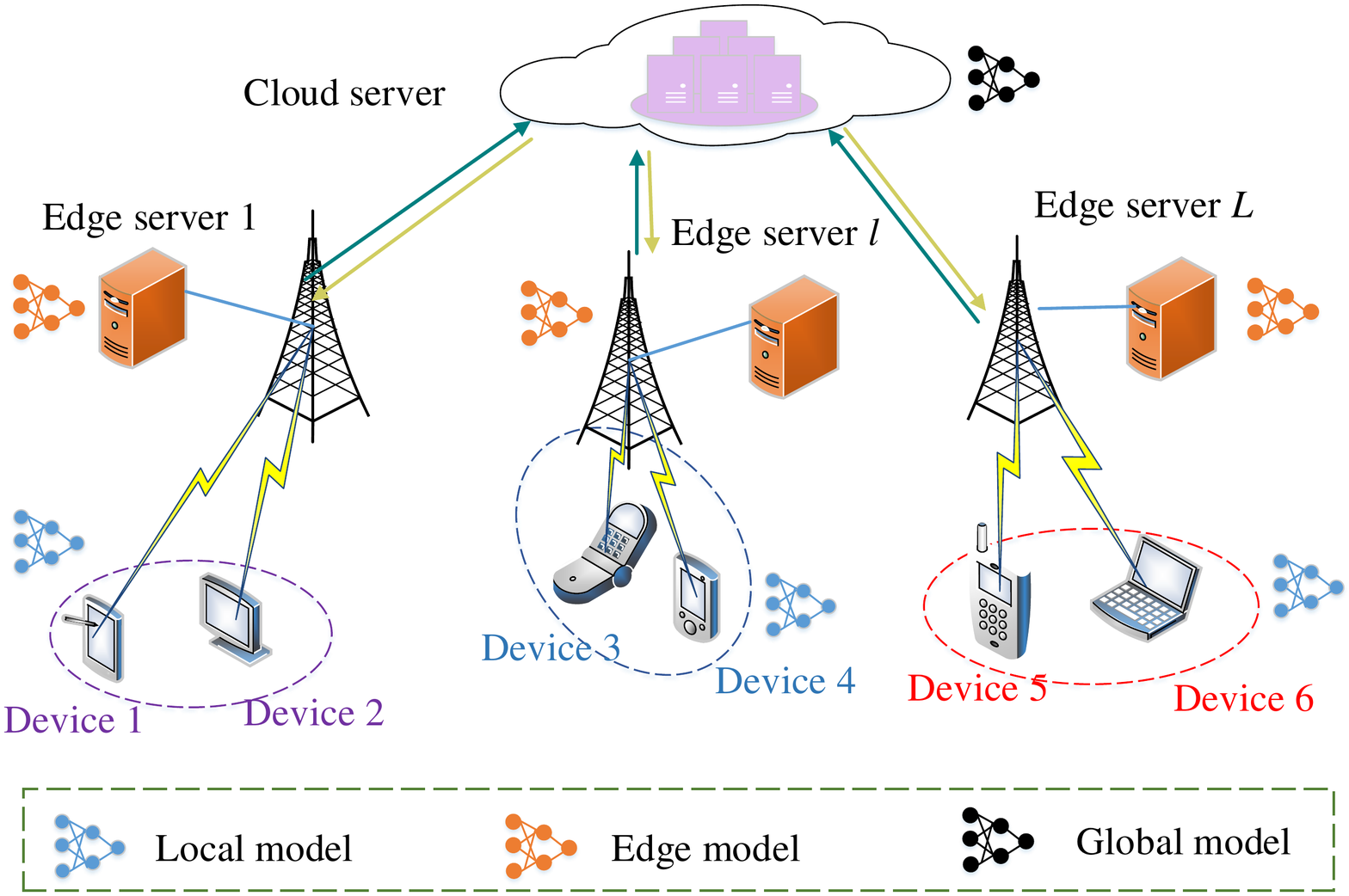}
\caption{An HFL system with a two-tiered computing structure, consisting of multiple devices and multiple edge servers. In the HFL process, the devices first transmit their local models to edge servers for edge aggregation, and then the edge servers send the edge models to the cloud server for global aggregation.}
\label{System_model}
\end{figure}
\section{System Model}
\label{sec:System Model}
Considering a wireless HFL system that consists of a set $\boldsymbol{\mathcal{N}}= \{1,...,n,...,N\}$ devices. There exists a distinct cloud server, and $L$ edge servers, e.g., base stations, employed across the network to aggregate the local models from the devices \cite{joint2020,dynamic2021}, the set of which is denoted
by $\boldsymbol{\mathcal{L}}=\{1,...,l,...,L\}$, as shown in Fig. \ref{System_model}. Overlap exists in the coverage of edge servers, and devices located within these areas often have multiple candidate edge servers available for connection. We assumes quasi-static Rayleigh fading for the communication channels, meaning that the channel coefficients between the devices and the edge servers are considered to be constant over a specific period of time. Moreover, we assume that the system bandwidth, denoted by $B$ MHz, can be partitioned into several subchannels for transmitting model parameters.  These subchannels are managed by all the edge servers participating in the HFL system.  Dividing the total bandwidth into subchannels allows multiple devices to send their model parameters concurrently without causing interference. In the HFL process, each device is able to select an edge server to associate with. Once the association is established, the device will transmit its updated parameters to its associated edge server for edge aggregation. After multiple rounds of edge aggregation, the edge servers send the aggregated edge models to the cloud server for global model aggregation. The specific steps of HFL have been described as follows.
\subsection{Hierarchical Federated Learning Model}
In the HFL system, the task for the coterie is to leverage the local datasets of all devices without jeopardizing their privacy to train a general supervised machine learning model, such as an SVM, logistic regression, or neural network. The training process of HFL can be divided into three steps: device updates, edge aggregation, and cloud aggregation. For clarity, we refer to the model of aggregated by the cloud server as the global model, the model aggregated by the edge servers as the edge model, and the model trained by each device using its local dataset as the local model. Thus, HFL is an iterative approach as each global aggregation includes multiple rounds of edge aggregations and each edge aggregation involves multiple rounds of local updates.

\textbf{Device Updates.} In each communication round, each device $n \in \{1,...,n,...,N\}$ first downloads the edge model from the edge server $l$ denoted by $\omega_l^{(i)}$ and trains the model locally with its own dataset. Then, each device $n$ transmits its updated local model $\omega_{n,l}^{(i)}$ to its associated edge server $l$. Here, we define the local training time of device $n$ as $T^{\text{cop},l}_n$ and the upload time of its local model as $T^{\text{com},l}_{n}$. It should be noted that the broadcast channel bandwidth of each edge server is sufficiently wide and its transmission power is much higher than that of the devices. Therefore, we can ignore the download time of the edge model for each device without any loss of generality.

\textbf{Edge Aggregation.} Upon receiving the local models from the devices, the edge server $l$ performs aggregation of the local parameters from its associated devices, e.g. FedAvg, to obtain an updated edge model $\omega_l^{(i+1)}$, which is transmitted back to the devices for the $(i+1)^{\text{th}}$ iteration. The immense computational capabilities of edge servers render the time required for local model aggregation negligible in most cases. We define he execution time for an edge aggregation of edge server $l$ as $T_{l}^{\text{edge}}$, which incorporates the time required by devices to perform local updates and upload their models.

\textbf{Cloud Aggregation.} Each edge server transmits the edge model parameters to the cloud sever for global aggregation to derive the updated global model. Here, we assume that the time interval between two global aggregations is fixed and denoted by $T^{\text{cloud}}$. Furthermore, we assume that the number of edge aggregations of the edge server $l$ in a global aggregation is $K_l$ and the index of the edge aggregations is $i \in \{0,...,K_l-1\}$. Since the connections linking edge servers and cloud servers are wired,  the communication time required for transmitting the model between them can be ignored.

Let $\boldsymbol{H}=[h_{n,l}]_{N\times L}$ be a channel gain matrix where $h_{n,l}$ is the channel gain between device $n$ and edge server $l$. Before participating in the HFL system, each device will evaluate different edge servers based on their own channel conditions and computing capability, and decide which edge server to associate with. We also define $\mathcal{S}_l$ as the set of devices that provide local training to edge server $l$, i.e., $\mathcal{S}_l = \{n \in \boldsymbol{\mathcal{N}} : a_n= l\}$, which satisfies $\mathcal{S}_l \subseteq \boldsymbol{\mathcal{N}}$ and $\cup_{l \in \boldsymbol{\mathcal{L}}}\mathcal{S}_l= \boldsymbol{\mathcal{N}}$. $a_n$ is the coalition selection of device $n$, which can be defined as an action vector $\boldsymbol{A}=[a_n]_{1 \times N}$. Here, we regard all the devices $S_l$ serving the edge server $l$ as a coalition, and we define $\boldsymbol{\mathcal{S}}$ as a coalition partition.

Due to the synchronous model aggregation mechanism, the execution time for an edge aggregation of edge server $l$ is determined by the number $K_l$ of edge aggregation performed by edge server $l$ within the time interval $T^{\text{cloud}}$. Thus, the execution time for an edge aggregation of edge server $l$ is described as
\begin{equation}
\begin{split}
\label{time for intermediate aggregation}
T_l^{\text{edge}}=\frac{T^{\text{cloud}}}{K_l}.
\end{split}
\end{equation}

After receiving the edge model from the edge server, the device performs local training with its own data samples to update its local model. For the device $n$ in coalition $\mathcal{S}_{l}$, its local training time can be expressed as
\begin{equation}
\begin{split}
\label{time for local training}
T^{\text{cop},l}_n=\frac{D_{n}^l C_n}{f_n},
\end{split}
\end{equation}
where $D_n^l$ means the total data quantity of device $n$ in coalition $l$ for local training, $f_n$  is regarded as a measurement of computation capacity of device $n$, that is, the CPU frequency, and $C_n$ is the number of CPU cycles to train a unit sampled data on device $n$.
Upon completing local training, every device transmits its trained local model to the corresponding edge server to perform edge aggregation. Following \cite{dynamic2021}, each device $n$ in the coalition $\mathcal{S}_{l}$
is equally allocated the wireless bandwidth resources. This assumption is appropriate as all devices are participating towards training the same global model. Thus, the uplink
rate the device $n$ is given as follows:
\begin{equation}
\begin{split}
\label{rate for uplink}
R^{\text{com},l}_{n}=\frac{B_l}{|\mathcal{S}_{l}|} \log(1+\frac{P_n h_{n,l}}{\sigma^2}),
\end{split}
\end{equation}
where $B_l$ is the bandwidth allocation for coalition $\mathcal{S}_{l}$ which satisfies $\sum_{l=1}^L B_l \le B$, $B_l \geq 0$. $|\mathcal{S}_{l}|$ indicates the number of devices in coalition $\mathcal{S}_{l}$. In addition, $P_n$ refers to the transmit power of the device $n$ and $\sigma^2$ is the power of the additive white Gaussian noise. We assume that the size of the local model of all devices is uniform and denoted by $s$. Therefore, the uplink time for the local parameters of device $n$ in coalition $\mathcal{S}_{l}$ can be given by
\begin{equation}
\begin{split}
\label{time for uplink}
T^{\text{com},l}_{n}=\frac{s}{R_{n}^{\text{com},l}}.
\end{split}
\end{equation}

\subsection{Participant Utility Function of Edge FL}
In general, all edge servers aim to obtain a well-trained edge model with a low loss function (or high  model accuracy). Given a fixed number of edge aggregation $K_l$ of the edge server $l$, the upper bound of expected difference $F(\omega_l^{K_l})-F(\omega_l^*)$ is increasing and convex function with respect to the amount of data from all devices that serve the edge server $l$, and satisfies a diminishing marginal effect. According to \cite{ren2021acclerating,huang2022collaboration}, after $K_l$ edge aggregations, the model improvement $\Lambda_l$ is dominated by term $\sqrt{K_l\sum_{n \in \mathcal{S}_{l}} D_n^l}$ and can
be approximately expressed as:
\begin{equation}
\begin{split}
\label{model improvement_dtat}
\Lambda_l=\xi \sqrt{K_l\sum_{n \in \mathcal{S}_{l}} D_n^l},
\end{split}
\end{equation}
where $\xi$ is a coefficient determined by the specific neural network model, and $\sqrt{\sum_{n \in \mathcal{S}_{l}} D_n^l}$ refers to the amount of data of all devices in coalition $\mathcal{S}_l$. Therefore, the training
revenue of coalition $\mathcal{S}_l$ paid by the edge server $l$, i.e., $r_{\mathcal{S}_l}$, is given by

\begin{equation}
\begin{split}
\label{traing revenue}
r_{\mathcal{S}_l}=\rho_l \xi \sqrt{K_l \sum_{n \in \mathcal{S}_{l}} D_n^l}+ |\mathcal{S}_{l}| x_l,
\end{split}
\end{equation}
where $\rho_l$ represents the unit price provided by the edge server $l$ for the model improvement brought on by the coalition $\mathcal{S}_{l}$'s local training. Since each edge server has different preferences for the improvement of edge FL performance, the unit price $\rho_l$ is also different between different edge servers. And $x_l$ is a fixed reward offered to devices in coalition $l$ based on the compensation for the devices participation costs.

In addition, the limited wireless capacity in the physical communication may lead to the congestion effect, which in turn leads to an increase in edge FL training latency. Specifically, with more devices joining a particular coalition, the congestion effect is greater, thereby resulting in cost among the devices. The cost of the device $n$ from joining the coalition $\mathcal{S}_l$ that results in the congestion
effect \cite{Shyuan2022reputation,gong2017when} is modelled as follows:
\begin{equation}
\begin{split}
\label{congestion cost}
z_n^{\mathcal{S}_l} = \alpha_l \left( \sum_{n \in \mathcal{S}_{l}} R^{\text{com},l}_{n}\right)^2,
\end{split}
\end{equation}
where $\alpha_l>0$ is the congestion coefficient that is determined by the resource constraint of edge server $l$ and $\sum_{n \in \mathcal{S}_{l}} R^{\text{com},l}_{n}$ is the total usage of communication resources of
edge FL devices in coalition $\mathcal{S}_l$. Thus, the utility of all devices in coalition $\mathcal{S}_l$, denoted by $u^{\mathcal{S}_l}$, can be expressed as follows.
\begin{equation}
\begin{split}
\label{coalition utility}
u^{\mathcal{S}_l}&= r_{\mathcal{S}_l} - |\mathcal{S}_{l}| z_n^{\mathcal{S}_l}
= \rho_l \xi \sqrt{ K_l \sum_{n \in \mathcal{S}_{l}} D_n^l}+ |\mathcal{S}_{l}| x_l- |\mathcal{S}_{l}|  \alpha_l \left( \sum_{n \in \mathcal{S}_{l}} R^{\text{com},l}_{n}\right)^2.
\end{split}
\end{equation}
We also define the total utilities of all coalitions, i.e., the utility of all devices in edge FL network, as
\begin{equation}
\begin{split}
\label{system coalition utility}
u_{\mathcal{N}}= \sum_{\mathcal{S}_{l} \in \mathcal{N}} u^{\mathcal{S}_l}=  \sum_{\mathcal{S}_{l} \in \mathcal{N}} r_{\mathcal{S}_l} - |\mathcal{S}_{l}| z_n^{\mathcal{S}_l}.
\end{split}
\end{equation}
Thus, the optimization objective is to maximize the total utility of all devices in edge FL network under bandwidth and delay constraints for all coalition, which can be described as follows.

\begin{align}
\mathop{\max}_{\boldsymbol{A},\boldsymbol{B},\boldsymbol{D}}\ \qquad &u_{\mathcal{N}}，\label{maximize the utility of all devices}\\
\quad\text{s.t.}\qquad
&\sum_{l=1}^L B_l \leq B,
 \tag{\ref{maximize the utility of all devices}{a}}\label{constraint1 a}\\
& T^{\text{cop},l}_n + T^{\text{com},l}_{n} \leq T_l^{\text{edge}}, \tag{\ref{maximize the utility of all devices}{b}}\label{constraint1 b}\\
&B_l \geq 0,\quad\quad \forall n, \forall l. \notag
\end{align}
(\ref{constraint1 a}) indicates that the bandwidth allocation for all coalitions must not exceed the total bandwidth $B$, and (\ref{constraint1 b}) means that the time of local training and model upload for each device in a coalition must be less than the execution time of a single edge aggregation in that coalition. Additionally, it should be noted that $\boldsymbol{A}$, $\boldsymbol{B}$, and $\boldsymbol{D}$ correspond to the action vector for all devices, the bandwidth allocation vector for coalitions, and the vector of the amount of data for local training by devices, respectively, i.e., $\boldsymbol{A}=[a_n]_{1 \times N}$, $\boldsymbol{B}=[B_l]_{1 \times L}$, and $\boldsymbol{D}=[D_n^{a_n}]_{1\times N}$.

\section{Low-Level Coalition Formation Game Formulation}
\label{sec:Coalition formation game formulation}
\subsection{Game Model and Analysis}
Coalition game is an excellent tool for revealing the coalition formulation process. Through the utilization of the coalition game in the edge FL network, devices can collaborate to form coalitions that improve the efficiency of the edge FL network. Thus, the problem of maximizing the total utility of all devices in edge FL network can be modeled as a coalition formulation game with transferable utility, in which the entire utility of the coalition can be distributed among the devices in any way in the coalition. In order to ensure active participation by the devices in the edge FL, the utility of the coalition can be distributed among the players within the coalition using an appropriate fairness rule.

Motivated by \cite{Lim2022decentralized}, we adopt the equal fair allocation rule. The utility derived by device $n$ from the coalition $l$ is given by
\begin{equation}
\begin{split}
\label{device j m reward2}
u_{n}^{\mathcal{S}_l}=\left(\rho_l \xi \sqrt{K_l \sum_{i \in \mathcal{S}_l}D_{i}^l}\right)\frac{D_{n}^l }{\sum_{i \in \mathcal{S}_l}D_{i}^l}+x_l-\alpha_l \left( \sum_{n \in \mathcal{S}_{l}} R^{\text{com},l}_{n}\right)^2,
\end{split}
\end{equation}
where $\frac{D_{n}^l }{\sum_{i \in \mathcal{S}_l}D_{i}^l}$ means the share of rewards based on data contribution across devices. According to (\ref{device j m reward2}), we know that each device can enhance its own utility by adjusting the amount of data for training $D_{n}^l$. Thus, it is necessary to determine the optimal amount of training data for each device within the coalition in order to maximize its utility. Prior to presenting the otimal strategy for the amount of data on the device, we will introduce a fundamental definition that is commonly used in coalition game.

\begin{myDef}
	\label{definition 1}
    \textit{(Coalition partition)} The set $\boldsymbol{\mathcal{S}} = \{ \mathcal{S}_1, ...,\mathcal{S}_L \}$ is a coalition partition of $\mathcal{N}$ if $\mathcal{S}_l \cap \mathcal{S}_i=\emptyset$, $\forall l, i \in \mathcal{L}, i\neq l$ and $\cup_{i \in \mathcal{L}} \mathcal{S}_i = \mathcal{N}$.
\end{myDef}

An example of the coalition partition has been depicted in Fig.~\ref{System_model}, where $\mathcal{N}=\{1, 2, 3, 4, 5, 6\}$, $\mathcal{S}_1=\{1, 2\}$, $\mathcal{S}_l=\{3, 4\}$, and $\mathcal{S}_L=\{5, 6\}$. Therefore, the set $\mathcal{S}=\{\{1, 2\}, \{3, 4\}, \{5, 6\}\}$ is a coalition partition of $\mathcal{N}$.

According to (\ref{device j m reward2}), the utilities of devices in the same coalition will influence each other for a given coalition partition. In other words, the utility of a device is influenced by not only its own strategy but also the strategies adopted by the other devices in the coalition regarding the amount of local training data. As a result, there exists a competition game between devices in the same coalition. Each device in the coalition intends to maximize its own utility, we need to find a Nash equilibrium profile that can be accepted by all participants in the competition game.

\begin{theorem}
	\label{theorem 1}
   Given a coalition partition $\boldsymbol{\mathcal{S}} = \{ \mathcal{S}_1, ...,\mathcal{S}_L \}$, the Nash equilibrium strategy for the amount of local training data of the device $i \in \mathcal{S}_l$ in the coalition $l \in \{1, ..., L\}$ is
\begin{equation}
\begin{split}
\label{Nash equilibrium strategy}
D_{i}^{l*}&=\left(T_l^I-\frac{s}{R^{\text{com},l}_{i}}\right)\frac{f_i}{C_i}
=\left(\frac{\Delta T}{K_l}-\frac{s |\mathcal{S}_l|}{B_l \log (1+\frac{P_i h_{i,l}}{\sigma^2})} \right)\frac{f_i}{C_i},
\end{split}
\end{equation}
and the Nash equilibrium strategy profile of the coalition $l$ is $\boldsymbol{D^{l*}}= \{D_{1}^{l*}, ..., D_{i}^{l*}, ..., D_{|\mathcal{S}_l|}^{l*}\}$, $\forall l \in \{1, ..., L\}$.

\emph{Proof:}\quad We can easily determine from (\ref{device j m reward2}) that $\left(\rho_l \xi \sqrt{K_l \sum_{j \in \mathcal{S}_l}D_{j}^l}\right)\frac{D_{i}^l }{\sum_{j \in \mathcal{S}_l}D_{j}^l}$ is an increasing function with respect to $D_{i}^l$. It means that the more data the device $i$ uses for local training, the more rewards it will receive. Besides, each device need to meet the delay constraints, i.e., (\ref{constraint1 b}). As a result, the maximum amount of data for local training taken by the device $i$ is the bounded value of the constraint (\ref{constraint1 b}). The noncooperative game can reach Nash equilibrium if and only if none of devices in a certain coalition can unilaterally modify the strategy to improve its utility. We assume that the strategy profile $\boldsymbol{D^{l*}}= \{D_{1}^{l*}, ..., \widetilde{D}_{i}^{l}, ..., D_{|\mathcal{S}_l|}^{l*}\}$ is the Nash equilibrium of the coalition $l$, where $\widetilde{D}_{i}^{l}$ indicates any non-negative real number less than $D_{i}^{l*}$. According to (\ref{device j m reward2}), we know that the strategy $\widetilde{D}_{i}^{l}$ is not the best response to the device $i$, and the device $i$ will modify its current strategy to $D_{i}^{l*}$, which is contrary to the definition of Nash equilibrium. Thus, the strategy profile $\boldsymbol{D^{l*}}= \{D_{1}^{l*}, ..., D_{i}^{l*}, ..., D_{|\mathcal{S}_l|}^{l*}\}$, $\forall l \in \{1, ..., L\}$ is the Nash equilibrium of the coalition $l$. $\hfill\blacksquare$
\end{theorem}

All devices are allowed to associate to the any edge server that meets the latency constraints, and resulting in the formation of coalitions. However, the rewards assigned to the devices usually vary from coalition to coalition. As a result, devices have different preferences over different coalitions. The preferences of devices over different coalitions are defined as follows.

\begin{myDef}
	\label{definition 2}
    \textit{(Preference relation)} For any device $n \in \mathcal{N}$, the preference relation or order $\succeq_n$, can be defined as a complete, reflexive, and transitive binary relation or order over the set of all coalitions that device $n$ could be a part of.
\end{myDef}

For a device $\forall n \in \mathcal{N}$, it decides to abandon or associate an edge server, i.e., leave or join the coalition. For example, for a device $n$, given two disjoint coalitions $\mathcal{S}_1 \subseteq \mathcal{N}$ and $\mathcal{S}_2 \subseteq \mathcal{N}$ , $\mathcal{S}_1 \succeq_n \mathcal{S}_2$ means device $n$ prefers being a member of coalition $\mathcal{S}_1$ rather than $\mathcal{S}_2$, or at least, device $n$ prefers $\mathcal{S}_1$ and $\mathcal{S}_2$ equally. Besides, $\mathcal{S}_1 \succ_n \mathcal{S}_2$ indicates that device $n$ strictly prefers to be a member of coalition $\mathcal{S}_1$ compared to coalition $\mathcal{S}_2$.

Given the finite set of all devices $\mathcal{N}$ and the preference order $\succ_n$ of any device $n$, the coalition formation game is thus defined as follows.
\begin{myDef}
	\label{definition 3}
    The proposed game is a coalition formation game which can be defined by the pair $(\mathcal{N},\succeq)$, where $\mathcal{N}$ is the finite set of devices in the HFL system and $\succeq$ is the preferences profile defined for each device in $\mathcal{N}$.
\end{myDef}

According to previous work \cite{Hao2012hedonic,Chen2021joint}, we know that the utility of each device, the structure of the coalitions and the game's convergance are all affected by the preference rule.
Here, we will define some typical preference rules as follows.

\begin{myDef}
	\label{definition 4}
    (Selfish preference rule) If there are any two potential coalitions that can be joined for device $n$, $\forall n \in \mathcal{N}$, i.e., $\mathcal{S}_l \subseteq \mathcal{N}$ and $\mathcal{S}_j \subseteq \mathcal{N}$, then the preference relation is
\begin{equation}
\begin{split}
\label{selfish preference rule definition}
\mathcal{S}_l \succeq_n \mathcal{S}_j \Leftrightarrow u_{n}^{\mathcal{S}_l} \geq u_{n}^{\mathcal{S}_j}.
\end{split}
\end{equation}
\end{myDef}

According to \emph{Definition \ref{definition 4}}, we know that each device in a coalition only thinks about its own utility and disregards the utilities of the other devices.
Each device may frequently move from the original coalition to a new coalition in the pursuit of high utility, which will hurt the interests of other devices. To address this quandary, a preference rule that considers the coalition's utilities has been proposed. As a typical preference rule, Pareto preference can gradually improve the utility of the device while preserving the others utilities in the original and new coalitions. Thus, the Pareto preference rule is defined as follows.
\begin{myDef}
	\label{definition 5}
    (Pareto preference rule) If there are any two potential coalitions that can be joined for device $n$, $\forall n \in \mathcal{N}$, i.e., $\mathcal{S}_l \subseteq \mathcal{N}$ and $\mathcal{S}_j \subseteq \mathcal{N}$, $l \neq j$, then the preference relation is
\begin{equation}
\begin{split}
\label{Pareto preference rule definition}
\mathcal{S}_l \succeq_n \mathcal{S}_j \Leftrightarrow u_{n}^{\mathcal{S}_l} \geq u_{n}^{\mathcal{S}_j} \wedge u_i^{\mathcal{S}_l} \geq u_i^{\mathcal{S}_l\setminus \{n\}},\\ \forall i \in \mathcal{S}_l\setminus \{n\}
\wedge u_i^{\mathcal{S}_j} \leq u_i^{\mathcal{S}_j\setminus \{n\}}, \forall i \in \mathcal{S}_j\setminus \{n\}.
\end{split}
\end{equation}
\end{myDef}

From (\ref{Pareto preference rule definition}), we can see that the Pareto preference rule have  stronger restrictions than the selfish preference rule. In such rule, as the device $n$'s utility is increased, those of others in the original and new coalitions are also increased. This implies that the device $n$ improves its own utility while causing no harm to the other devices in the original and new coalitions. However, the strong restrictions in the Pareto preference rule may make it difficult for devices to leave the original coalition and join the new one, making it challenging to improve both the utilities of the devices and the coalitions. A new preference rule that takes into account the total utilities of the original and new coalitions is suggested as a solution to the aforementioned issue. Thus, we define the coalition altruistic preference rule as follows.

\begin{myDef}
	\label{definition 6}
    (Coalition altruistic preference rule) If there are any two potential coalitions that can be joined for device $n$, $\forall n \in \mathcal{N}$, i.e., $\mathcal{S}_l \subseteq \mathcal{N}$ and $\mathcal{S}_j \subseteq \mathcal{N}$, then the preference relation is
\begin{equation}
\begin{split}
\label{coalition altruistic preference rule}
\mathcal{S}_l \succeq_n \mathcal{S}_j \Leftrightarrow \sum_{i \in \mathcal{S}_l } u_i^{\mathcal{S}_l} + \sum_{i \in \mathcal{S}_j \setminus \{ n \}} u_i^{\mathcal{S}_j \setminus \{ n \}}\\ \geq  \sum_{i \in \mathcal{S}_l \setminus \{ n \}} u_i^{\mathcal{S}_l \setminus \{ n \}} + \sum_{i \in \mathcal{S}_j} u_i^{\mathcal{S}_j}.
\end{split}
\end{equation}
\end{myDef}
It is clear from (\ref{coalition altruistic preference rule}) that device $n$ will joined the coalition which could increase the total utility of itself and other devices in both original and new coalition.

It is obvious that the device $n$ can join a new coalition once the aforementioned preference rules are satisfied. Typically, the utility of each device is determined by the edge server to which it is connected and the bandwidth allocated to the coalition. The bandwidth resources allocated to each device often influence its model upload latency, which in turn can affect its local training performance. Therefore, given a coalition partition, the corresponding bandwidth allocation rule is provided as follows.

\begin{myDef}
	\label{definition 7}
    (Bandwidth reallocation rule) Given a coalition partition $\boldsymbol{\mathcal{S}} = \{ \mathcal{S}_1, ...,\mathcal{S}_L \}$, the device $i$, $i \in \mathcal{S}_l$ moves from the original coalition $\mathcal{S}_l$ to a new coalition $\mathcal{S}_j$, hence the total bandwidth owned by the original coalition $\mathcal{S}_l$ and the new coalition $\mathcal{S}_j$ will be reallocated. Besides, the bandwidth of other coalitions remains unchanged.
\end{myDef}

The bandwidth allocated to other coalitions remains unchanged in the proposed bandwidth reallocation rules because the members of those coalitions have not changed. Whereas, the bandwidth of the original and new coalition of the device $n$ will change as it changes. Now, we present the definition of the switch rule as follows.

\begin{myDef}
	\label{definition 8}
    (Switch rule) Given a coalition partition $\boldsymbol{\mathcal{S}} = \{ \mathcal{S}_1, ...,\mathcal{S}_L \}$, the device $i \in \mathcal{S}_l$, decides to move from the original coalition $\mathcal{S}_l$ to a new coalition $\mathcal{S}_j$, $j \neq l$, if and only if $\mathcal{S}_j \cup \{i\} \succeq_n \mathcal{S}_l$. Then, the new coalition partition can be described as $\widehat{\boldsymbol{\mathcal{S}}}=\{\boldsymbol{\mathcal{S}}\setminus \{\mathcal{S}_j, \mathcal{S}_l\},\mathcal{S}_l \setminus \{i\}, \mathcal{S}_j \cup \{i\}\}$.
\end{myDef}

The switch rule provides a mechanism through which the device $n$ can leave the original coalition and selected a new coalition to join if and only if the new coalition is preferred over the current coalition, which have been defined in (\ref{selfish preference rule definition}), (\ref{Pareto preference rule definition}) and (\ref{coalition altruistic preference rule}). Next, we define the stable partition in our coalition formation game.

\begin{myDef}
	\label{definition 9}
    (Stable partition) A coalition partition $\boldsymbol{\mathcal{S}} = \{ \mathcal{S}_1, ...,\mathcal{S}_L \}$ is stable if there is no device $n$ with a preference for other coalitions. In other words, there is no device tends to join a new coalition if the stability coalition partition of the coalition game has formed.
\end{myDef}
We can obtain the following theorem through a simple proof.

\begin{theorem}
	\label{theorem 2}
  Under selfish and Pareto orders, the coalition partition of our coalition game can eventually converge to a stable coalition partition.

\emph{Proof:}\quad We denote $\boldsymbol{\mathcal{S}}^{(0)}$ and $\boldsymbol{\mathcal{S}}^{(F)}$ as the initial coalition partition and the terminal stable coalition partition after finite switch operation separately. Thus, the switch sequence can be described as $\boldsymbol{\mathcal{S}}^{(0)} \rightarrow \cdots \rightarrow \boldsymbol{\mathcal{S}}^{(F)}$. Now, we use the proof by contradiction to demonstrate \emph{Theorem} \ref{theorem 2}. We assume that the final partition $\boldsymbol{\mathcal{S}}^{(F)}$ is not stable, i.e., there exist a device need to execute switch operation. In Pareto rule, the device will join a new coalition which satisfies the preference rule (\ref{Pareto preference rule definition}). Similarly, to pursue high personal utility in (\ref{selfish preference rule definition}), the device will switch its coalition in selfish rule. $\boldsymbol{\mathcal{S}}^{(F)}$ is not a final coalition partition which is contradicted with the previous argument. Thus, the final coalition partition is stable, and \emph{Theorem} \ref{theorem 2} has been proved.
 $\hfill\blacksquare$
\end{theorem}

Although Selfish and Pareto preference rules can converge to a stable partition, they are not considered in terms of global optimization, which may not ensure that a global optimal solution is achieved. While the coalition altruistic preference rule is considered from a coalition standpoint, which can be viewed as a partially collaborative approach. As a result, it is critical to investigate the stability under the coalition altruistic preference rule.

\begin{myDef}
	\label{definition 10}
    (Exact potential function) A game is an exact potential game when the potential function $\psi: \boldsymbol{\mathcal{A}}_1 \times \cdots \times \boldsymbol{\mathcal{A}}_N \rightarrow \mathcal{R}$ satisfies the following equation for $\forall n \in \mathcal{N}$:
\begin{equation}
\begin{split}
\label{Exact potential game definition}
&U_n(\widetilde{a}_n, a_{-n})- U_n(a_n, a_{-n}) = \psi(\widetilde{a}_n, a_{-n})- \psi(a_n, a_{-n}), \
\forall a_n, \widetilde{a}_n \in \boldsymbol{\mathcal{A}}_n,
\end{split}
\end{equation}
where $a_{-n}$ means the edge server selections of other devices except device $n$.
\end{myDef}

\begin{theorem}
	\label{theorem 3}
   Thus, the coalition altruistic preference rule in our coalition formulation game contains at least one pure Nash equilibrium of the coalition partition.

   \emph{Proof:} Refer to Appendix \ref{Appendix A}. $\hfill\blacksquare$
\end{theorem}

According to the preceding theorem, there is always a stable coalition partition under the coalition altruistic preference rule. The utilities of the switched coalitions continuously increase during the switch operation, which is similar to the evolutionary path of Nash equilibrium.
\subsection{Coalition Game Based Algorithm}
In this subsection, we will focus on the algorithm for forming an effective coalition partition with bandwidth allocation for each coalition. Each device can automatically form coalitions based on the switch operation under the preference rule, and the coalition partition of the introduced coalition game may change over iterations. Thus, we use a coalition formation algorithm for the edge-side FL in which each device is associated with an edge server to form disjoint coalitions and assign bandwidth to each coalition.

Now, we discuss the coalition formation algorithm which is executed by the devices. In each iteration, an arbitrary device is chosen
to make decision. First, the selected device randomly chooses another coalition to explore its expected utility. It is noted that the
bandwidth reallocation only occurs in the original and new coalition. After that, the selected device makes a comparative update under coalition expected altruistic preference rule to decide
whether to join another coalition or keep executing local training in the original coalition. When the coalitions are determined,
the total revenue of all coalitions is determined. Thus, the goal is to improve the total utility of devices by scheduling
bandwidth under current coalition partition. Once the bandwidth allocation and the numbers of edge aggregation are determined, it in turn affects the coalition
formation process. Thus, devices will continue to update the
coalition partition. The iteration repeats until the coalition
partition converges to a final stable partition where no device
tends to deviate from its current coalition. The concrete
procedure is given in step 2 to 4 of Algorithm 1.

\begin{algorithm}[h]
\label{Implementation of Hierarchical Game}
    \caption{Implementation of Two-Level Incentive Mechanisms}%
    \KwIn{Let devices randomly choose a edge server $l$ , $\forall l \in \mathcal{L}$ and form an initial partition $\mathcal{S}_0$. Input $\mathcal{S}_0$ to Algorithm 2 to obtain the bandwidth allocation of each edge server $B_0$. Input $\mathcal{S}_0$ and the bandwidth allocation to Algorithm 3 to obtain the numbers of edge aggregations for each coalition.}
    \KwOut{The stable coalition partition $\boldsymbol{\mathcal{S}}$, the bandwidth allocation $\boldsymbol{B}$, the number of edge aggregations $\boldsymbol{K}$, the unit reward vector $\boldsymbol{\chi}$.}
    \Repeat{\rm{the coalition partition converges to a final stable partition}}
    {Randomly select a device $n$, $\forall n \in \mathcal{N}$, with current coalition partition $\mathcal{S}_{\text{current}}$ ($\mathcal{S}_{\text{current}}=\mathcal{S}_0$  in the first iteration). The selected device measures its current utility $U_n(a_n)$;

    Device $n$ randomly selects another edge server $a'_n$. Input ${\mathcal{S}_{a_n} , \mathcal{S}_{a'_n}}$ to Algorithm 2 to reallocate bandwidth owned by coalition $\mathcal{S}_{a_n}$ and $\mathcal{S}_{a'_n}$ in coalition $\mathcal{S}_{a_n}$ and $\mathcal{S}_{a'_n}$. And then, the selected device measures its explored utility $U_n(a'_n)$ under this partition;

    If such switch operation from $\mathcal{S}_{a_n}$ to $\mathcal{S}_{a'_n}$, where $\mathcal{S}_{a'_n} \cup \{n\} \succeq_n  \mathcal{S}_{a_n}$, exists, perform the following steps:
    \begin{enumerate}
      \item  Leave the current coalition, i.e., $\mathcal{S}_{a_n} := \mathcal{S}_{a_n} \setminus \{ n \}$;
      \item Join the new coalition, i.e., $\mathcal{S}_{a'_n} := \mathcal{S}_{a'_n} \cup n$;
      \item Update the bandwidth reallocation and numbers of edge aggregations results for coalition $\mathcal{S}_{a_n}$ and $\mathcal{S}_{a'_n}$ while keep other tasks bandwidth allocation unchanged.
    \end{enumerate}

    The cloud server determines its reward for each edge server based on Algorithm 3. At the same time, the edge servers of the original coalition $\mathcal{S}_{a_n}$ and the new coalition $\mathcal{S}_{a'_n}$ determine the numbers of edge aggregations for all devices associated with them.
     }

\end{algorithm}

We then present a gradient projection based method for bandwidth allocation. When the coalition partition is determined, the rewards of all coalitions can be obtained easily. Thus, we design the following objective function:
\begin{equation}
\begin{split}
&\mathcal{G}(\boldsymbol{B})= \sum_{l \in \mathcal{L}} \left(\rho_l \xi \sqrt{ K_l \sum_{n \in \mathcal{S}_{l}} \left(\frac{\Delta T}{K_l}-\frac{s |\mathcal{S}_l|}{B_l \log (1+\frac{P_n h_{n,l}}{\sigma^2})} \right)\frac{f_n}{C_n}} \right.\\&\left.+ |\mathcal{S}_{l}| x_l- |\mathcal{S}_{l}|  \alpha_l \left( \sum_{n \in \mathcal{S}_{l}} \frac{B_l}{|\mathcal{S}_{l}|} \log(1+\frac{P_n h_{n,l}}{\sigma^2})\right)^2\right).
\end{split}
\end{equation}

Hence, the original problem can be expressed as
\begin{align}
\mathop{\max}_{\boldsymbol{B}}\ \qquad &\mathcal{G}(\boldsymbol{B})，\label{maximize the utility of all devices2}\\
\quad\text{s.t.}\qquad
&\sum_{l=1}^L B_l \leq B,
 \tag{\ref{maximize the utility of all devices2}{a}}\label{constraint1 a2}\\
& T^{\text{cop},l}_n + T^{\text{com},l}_{n} \leq T_l^{I}, \tag{\ref{maximize the utility of all devices2}{b}}\label{constraint1 b2}\\
&B_l \geq 0,\quad\quad \forall n, \forall l. \notag
\end{align}
Since the second derivative of $\mathcal{G}(\boldsymbol{B})$ with respect to $\boldsymbol{B}$ is less than 0, $\mathcal{G}(\boldsymbol{B})$ is a concave function with respect to $\boldsymbol{B}$, we can apply the GP based method to allocate bandwidth when the coalition partition is determined. The GP method is summarized in Algorithm 2. Through several iterations from step 2 to Step 4,
we can obtain the optimal bandwidth allocation of the problem
specified in (\ref{maximize the utility of all devices2}). The proof of convergence can be found in Theorem 3.4 in \cite{xiu2007A}.

\begin{algorithm}[h]
\label{Gradient Projection Based Method for Bandwidth Allocation}
    \caption{Gradient Projection Based Method for Bandwidth Allocation Under Determined Coalition Partition}%
    \KwIn{Set $k = 0$, and define the maximum iteration number $K^{\text{max}}$. Initialize $\boldsymbol{B(0)}$ and define the tolerance of accuracy $\varepsilon$.}
    \KwOut{The bandwidth allocation $\boldsymbol{B}$.}
    \Repeat{\rm{the objective value converges, or the maximum number of iterations is reached}}
    {Calculate the gradient $\bigtriangledown G(\boldsymbol{B})$;

    Calculate the projection
    \begin{equation}
    \boldsymbol{B}_{\text{{proj}}}= P_{\Omega_B}(\boldsymbol{B}+\bigtriangledown G(\boldsymbol{B}));
    \end{equation}

    Update B according to the following rule
    \begin{equation}
    \boldsymbol{B} \leftarrow \boldsymbol{B} + \gamma (\boldsymbol{B}_{\text{{proj}}}-\boldsymbol{B});
    \end{equation}

     }

\end{algorithm}

%
%
%
%
%

\section{Upper-Level Stackelberg Game}
\label{sec:Upper-Level Stackelberg Game}
In this section, we design a Stackelberg game algorithm to jointly maximize the utilities of the cloud and each edge server.

\subsection{Problem Formulation of Upper-Level Game}
At the beginning of the global aggregation, one device has the option to leave its original coalition and join a new one. The edge servers of the original coalition and the new coalition will determine the numbers of edge aggregations for all devices associated with them as a response. Therefore, given a coalition partition, the corresponding edge aggregations rule is provided as follows.
\begin{myDef}
	\label{edge aggregations rule}
    (Edge aggregations rule) Given a coalition partition $\boldsymbol{\mathcal{S}} = \{ \mathcal{S}_1, ...,\mathcal{S}_L \}$, the device $i$, $i \in \mathcal{S}_l$ moves from the original coalition $\mathcal{S}_l$ to a new coalition $\mathcal{S}_j$, the numbers of edge aggregations determined by the edge servers of the original coalition $\mathcal{S}_l$ and the new coalition $\mathcal{S}_j$ will be reallocated. Besides, the numbers of edge aggregations determined by the edge servers of other coalitions remains unchanged.
\end{myDef}

 The number of edge aggregations not only affects the performance of HFL but also impacts the utilities of edge servers and devices. The the edge servers maximize their utilities by adjusting the number of edge aggregations. Correspondingly, the cloud server reallocate rewards to the edge servers that modify their number of edge aggregations to maximize its utility and enhance the learning efficiency of the HFL system. To facilitate this interaction, we utilize a Stackelberg game approach.

Following \cite{li2014efficient} and \cite{ding2020optimal}, we know that the accuracy loss of the edge server $l$ after $K_l$-round edge aggregation is measured by the difference between the prediction loss with parameter $\omega_l^{K_l}$ and that with the optimal parameter $\omega_l^*$, whose expectation is bounded by $\frac{\lambda}{\sqrt{(K_l \sum_{i \in \mathcal{S}_l}D_{i}^l)}}+\frac{\lambda}{K_l}$. It is essential to formulate a utility function that takes into account the overall learning performance for the cloud server, who plays the role of the leader in the Stackelberg game. Based on \cite{mai2022automatic}, we define the utility function of the cloud server as (\ref{cloud reward_new}), where the first term represents the reward obtained by the cloud service due to the global improvement in HFL's performance, and the second term denotes the reward disbursed by the cloud server to all edge servers for edge aggregation.
\begin{figure*} 
 	\centering
 	\begin{equation}
    \begin{split}	
 		u_{\text{cloud}}= \mathbf{H}\left( \sum_{l=1}^L \left[G-\frac{\lambda}{\sqrt{K_l \sum_{n \in \mathcal{S}_{l}} \left(\frac{\Delta T}{K_l}-\frac{s |\mathcal{S}_l|}{B_l \log (1+\frac{P_n h_{n,l}}{\sigma^2})} \right)\frac{f_n}{C_n}}} -\frac{\lambda}{K_l}\right]\right)
\\-\sum_{l=1}^L \chi_l \left(G-\frac{\lambda}{\sqrt{K_l \sum_{n \in \mathcal{S}_{l}} \left(\frac{\Delta T}{K_l}-\frac{s |\mathcal{S}_l|}{B_l \log (1+\frac{P_n h_{n,l}}{\sigma^2})} \right)\frac{f_n}{C_n}}}-\frac{\lambda}{K_l}\right).
   \end{split}
 		\label{cloud reward_new}
 	\end{equation}

 \end{figure*}
The function $\mathbf{H}(\cdot)$ is a positive, increasing and concave function, and is described as follows:
\begin{equation}
\begin{split}
\label{cloud utility_new}
\mathbf{H}(\boldsymbol{x})=\beta(\frac{a}{L} \boldsymbol{x} +b)^{0.5}.
\end{split}
\end{equation}
$\lambda$ indicates the edge servers' valuation on accuracy loss, and $G$ means the finite (and possibly large) accuracy loss for all edge server when there is no data for training.  $\beta$ is system parameter, $a$ and $b$ are dynamical parameters respectively. $\chi_l$ means the unit reward provided by the cloud server to the edge server $l$ for its edge aggregation performance.

As the followers of the game, the edge server need to obtain an optimal strategy of edge aggregations in response to the reward scheme of the leader in Stackelberg game. The utility function of edge server $l$ is defined as
\begin{equation}
\begin{split}
\label{edge server l reward}
u_{\text{edge}}^l&= \chi_l \Bigg(G-\frac{\lambda}{\sqrt{K_l \sum_{n \in \mathcal{S}_{l}} \left(\frac{T^{\text{cloud}}}{K_l}-\frac{s |\mathcal{S}_l|}{B_l \log (1+\frac{P_n h_{n,l}}{\sigma^2})} \right)\frac{f_n}{C_n}}} -\frac{\lambda}{K_l}\Bigg)
-|\mathcal{S}_l|x_l\\&- \left(\rho_l \xi \sqrt{K_l \sum_{n \in \mathcal{S}_{l}} \left(\frac{T^{\text{cloud}}}{K_l}-\frac{s |\mathcal{S}_l|}{B_l \log (1+\frac{P_n h_{n,l}}{\sigma^2})} \right)\frac{f_n}{C_n}}\right).
\end{split}
\end{equation}

Hence, the edge server $l$ determines the number of edge aggregations, by maximizing the following optimization problem, i.e.,
\begin{align}
\label{edge server l reward optimization problrm}
\max_{K_l}\qquad  &u_{\text{edge}}^l\\
\text{s.t.}\qquad
&\left(\frac{T^{\text{cloud}}}{K_l}-\frac{s}{R^{\text{com},l}_{n}}\right)\frac{f_n}{C_n}\geq 0,\  \forall n \in \mathcal{S}_l,\tag{\ref{edge server l reward optimization problrm}{a}}\label{constraint3 a3}\\
&\quad K_l \in \mathbb{N}^+, \forall l \tag{\ref{edge server l reward optimization problrm}{b}}\label{constraint3 b3},
\end{align}
where (\ref{constraint3 a3}) means that the number of edge aggregations $K_l$ must be taken in such a way that the amount of local training data for each device is not less than 0. (\ref{constraint3 b3}) indicates that $K_l$ must be a positive integer.

Correspondingly, the cloud server decides the unit price of performance improvement of HFL offered to each edge server, so as to maximize the function in Eq. (\ref{cloud reward_new}), i.e.,
\begin{align}
\label{cloud server l reward optimization problrm}
\max_{\boldsymbol{\chi}}\qquad  u_{\text{cloud}}
\qquad \qquad \text{s.t.}\quad
\chi_l \geq 0, \forall l \in \mathcal{L}. \notag
\end{align}
For the purposes of analysis, we make $A_l= T^{\text{cloud}} \sum_{n\in \mathcal{S}_l} \frac{f_n}{C_n}$ and $F_l = s |\mathcal{S}_l| \sum_{n\in \mathcal{S}_l} \frac{f_n}{C_n \log (1+\frac{P_n h_{n,l}}{\sigma^2})}$. Therefore, the utility function $u_{\text{edge}}^l$ of edge server $l$ can be replaced by
\begin{equation}
\begin{split}
\label{replace utility function of edge server}
u_{\text{edge}}^l=\chi_l \left(G-\frac{\lambda}{\sqrt{A_l-\frac{K_l F_l}{B_l}}}-\frac{\lambda}{K_l}\right)
-|\mathcal{S}_l|x_l-\left(\rho_l \xi \sqrt{A_l-\frac{K_l F_l}{B_l}}\right).
\end{split}
\end{equation}

Similarly, the cloud server's utility function can be re-expressed as follows:
\begin{equation}
\begin{split}
\label{replace utility function of cloud server}
u_{\text{cloud}}= \mathbf{H}\left( \sum_{l=1}^L \left[G-\frac{\lambda}{\sqrt{A_l-\frac{K_l F_l}{B_l}}}-\frac{\lambda}{K_l}\right]\right)
-\sum_{l=1}^L \chi_l \left(G-\frac{\lambda}{\sqrt{A_l-\frac{K_l F_l}{B_l}}}-\frac{\lambda}{K_l}\right).
\end{split}
\end{equation}

\subsection{Stackelberg Game Approach}
In our proposed Stackelberg game, the cloud server and edge servers have the goal of maximizing their own revenues. 

Since $K_l$ is a positive integer, we relax it to $K_l \in \left[1, \min_{n \in \mathcal{S}_l} \left\{\frac{\Delta T R_{n}^{\text{com},l}}{s}\right\}\right]$ for the sake of analysis.
Taking the second derivative of the utility function $u_{\text{edge}}^l$ for edge server $l$ with respect to $K_l$, we can obtain:
\begin{equation}
\begin{split}
\label{equation11}
\frac{\partial^2 u_{\text{edge}}^l}{\partial K_l^2}=\frac{F_l^2 \rho_l \xi}{4B_l^2\left(A_l-\frac{F_l K_l}{B_l}\right)^{\frac{3}{2}}}-
\frac{3F_l^2 \chi_l\lambda}{4B_l^2\left(A_l-\frac{F_l K_l}{B_l}\right)^{\frac{5}{2}}}-\frac{2\chi_l\lambda}{K_l^3},
\end{split}
\end{equation}
which cannot be directly judged for its positivity or negativity, and thus for the edge server utility function's concavity. To investigate the optimality of Eq. (\ref{replace utility function of edge server}), we set $Z_l=\sqrt{A_l-\frac{F_l K_l}{B_l}}$. Then, we regain the utility function as follows:
\begin{equation}
\begin{split}
\label{equation12}
&u_{\text{edge}}^l(Z_l)=\chi_{l} \left(G-\frac{\lambda}{Z_l}-\frac{F_l \lambda}{B_l \left(A_l-Z_l^2\right)}\right)
-\xi \rho_l Z_l-|\mathcal{S}_l|x_l,\\
&Z_l \in \left[\min_{n \in \mathcal{S}_l} \left\{\sqrt{A_l-\frac{F_l}{B_l s}\Delta T R_{n}^{\text{com},l}}\right\}, \sqrt{A_l-\frac{F_l}{B_l}}\right].
\end{split}
\end{equation}
%
Let us compute the first derivative of the function $u_{\text{edge}}^l(Z_l)$ with respect to $Z_l$:
\begin{equation}
\begin{split}
\label{equation14}
\frac{\partial u_{\text{edge}}^l(Z_l)}{\partial Z_l}=\frac{\chi_l \lambda}{Z_l^2}-\xi \rho_l-\frac{2F_l \chi_l Z_l \lambda}{B_l (A_l- Z_l^2)^2}.
\end{split}
\end{equation}
Similarly, we take the second derivative of the function $u_{\text{edge}}^l(Z_l)$ with respect to $Z_l$, i.e.,
\begin{equation}
\begin{split}
\label{equation15}
\frac{\partial^2 u_{\text{edge}}^l(Z_l)}{\partial Z_l^2}=-\frac{2\chi_l \lambda}{Z_l^3}-\frac{2F_l \chi_l \lambda}{B_l (A_l- Z_l^2)^2}-\frac{8F_l \chi_l Z_l^2 \lambda}{B_l (A_l- Z_l^2)^3},
\end{split}
\end{equation}
where the second derivative of $u_{\text{edge}}^l(Z_l)$ with respect to $Z_l$ is less than or equal to zero, the utility function $u_{\text{edge}}^l(Z_l)$ is concave.

Setting the first derivative of $u_{\text{edge}}^l(Z_l)$ with respect to $Z_l$ be equal to $0$, i.e.,
\begin{equation}
\begin{split}
\label{first derivative equal 0}
\frac{\partial u_{\text{edge}}^l(Z_l)}{\partial Z_l}=\frac{\chi_l \lambda}{Z_l^2}-\xi \rho_l-\frac{2F_l \chi_l Z_l \lambda}{B_l (A_l- Z_l^2)^2}=0
\Leftrightarrow \frac{\lambda}{Z_l^2}-\frac{F_lZ_l \lambda}{B_l(A_l-Z_l^2)^2}= \frac{\xi \rho_l}{\chi_l},
\end{split}
\end{equation}
the utility function $u_{\text{edge}}^l$ can achieve the maximum value.
Due to the lack of the closed-form solution for the optimal edge aggregation strategies of the edge server $l$. To analyze the utility of the game's leader, we use the tools of variational inequality. The feasible region  of $\boldsymbol{\chi}$ is convex, compact, and closed subspace of a finite dimensional Euclidean space, and the mapping $\frac{\partial u_{\text{edge}}^l(Z_l)}{\partial Z_l}$ in (\ref{first derivative equal 0}) is continuous. Thus, the variational inequality in our game is solvable, in which the cloud server cyclically update its unit rewards for edge servers. At each iteration, the cloud server tries to increase or decrease its unit reward $\chi_l$ to maximize its own utility while keeping the unit reward provided for other edge servers, i.e., $\boldsymbol{\chi}_{-l}$ unchanged. The Stackelberg optimal strategies can be obtained from Algorithm 3. Each edge server can obtain the optimal edge aggregation number based on the best pricing of the cloud server and (\ref{first derivative equal 0}) and $Z_l=\sqrt{A_l-\frac{F_l K_l}{B_l}}$.

\begin{algorithm}[h]
\label{The Stackelberg optimal strategies}
    \caption{The Stackelberg Optimal Strategies for Cloud Server}%
    \KwIn{List all potential edge aggregation numbers for the edge servers participating in this game and calculate corresponding unit rewards according to (\ref{first derivative equal 0}), which can be listed as $[\chi_l^1,...,\chi_l^{K_{\text{max}}}]$; set the step size $\zeta$, iteration $t=1$, let $\chi_{\max}^l= \max \ [\chi_l^1,...,\chi_l^{K_{\text{max}}}]$, $\chi_{\min}^l= \min \ [\chi_l^1,...,\chi_l^{K_{\text{max}}}]$.}
    \KwOut{The unit reward vector $\boldsymbol{\chi}$.}
    \While{the cloud server changes the unit rewards for the edge servers}
     {
     \For{each edge server $l$ participating in the game}
     {Computing $\chi_l^{\text{cur}}= [(\chi_l)^{t-1}+ \zeta]^{\chi_{\max}^l}_{\chi_{\min}^l}$, and finding the interval that $\chi_l^{\text{cur}} \in [\chi_l^k, \chi_l^{k'}]$ in the list $[\chi_l^1,...,\chi_l^{K_{\text{max}}}]$, where $k$ and $k'$ are adjacent integers;

     \uIf {$\max \quad \{u_{\text{cloud}}(\chi_l^k, \boldsymbol{\chi}_{-l}), u_{\text{cloud}}(\chi_l^{k'}, \boldsymbol{\chi}_{-l})\} > u_{\text{cloud}}((\chi_l)^{t-1}, \boldsymbol{\chi}_{-l})$}{$(\chi_l)^{t}= \arg\max \quad \{u_{\text{cloud}}(\chi_l^k, \boldsymbol{\chi}_{-l}), u_{\text{cloud}}(\chi_l^{k'}, \boldsymbol{\chi}_{-l})\}$;}
        \uElse{
        Computing $\chi_l^{\text{cur}}= [(\chi_l)^{t-1}- \zeta]^{\chi_{\max}^l}_{\chi_{\min}^l}$, and finding the interval that $\chi_l^{\text{cur}} \in [\chi_l^k, \chi_l^{k'}]$ in the list;

       \uIf {$\max \quad \{u_{\text{cloud}}(\chi_l^k, \boldsymbol{\chi}_{-l}), u_{\text{cloud}}(\chi_l^{k'}, \boldsymbol{\chi}_{-l})\} > u_{\text{cloud}}((\chi_l)^{t-1}, \boldsymbol{\chi}_{-l})$}{$(\chi_l)^{t}= \arg\max \ \{u_{\text{cloud}}(\chi_l^k, \boldsymbol{\chi}_{-l}), u_{\text{cloud}}(\chi_l^{k'}, \boldsymbol{\chi}_{-l})\}$;}
       \uElse{$(\chi_l)^{t}= (\chi_l)^{t-1}$;}
        \textbf{end}
        }
        \textbf{end}
    %
%
     }
     Set $(\boldsymbol{\chi}_{-l})^{t}=(\boldsymbol{\chi}_{-l})^{t-1}$;

     Set $t= t+1$;
     }
\end{algorithm}

\section{Simulation Results and Discussion}
\label{Simulation Results and Discussion}
We present the performance evaluation of two-level incentive mechanisms in HFL in this section. Considering all devices and edge servers are randomly distributed in a $1$ km $\times$ $1$ km fixed region. We set the wireless resource bandwidth in our system as $5$ MHz. The number of CPU cycles $C_n$ for each device to
perform local model training of unit data sampling takes range from $3\times10^9$cycle/unit. The simulation parameters that we use are as follows in Table I.

\begin{table}[h]
\caption{Simulation Parameter Settings}
\label{Parameters in Simulations}
\center
\begin{tabular}{|p{5cm} |p{2.6cm}|p{5cm}|p{2.0cm}|}
\hline
Nations & Values & Nations & Values \\
\hline
The number of devices $N$ & $6 \sim 18$ &The size of local parameter $s$ & $3\times10^6 ~\text{bit}$\\
\hline
The number of edge servers $L$ & $4$ &The congestion coefficient $\alpha_l$ & $[0.05,0.15]$\\
\hline
Total bandwidth $W$ & $5~\text{MHz}$ &The transmit power of device $P_n$ & $[0.2,0.5]~W$ \\
\hline
The number of CPU cycles $C$ & $3\times10^{9}$ $\text{cycles/unit}$ &The additive white Gaussian noise power $\sigma^2$ & $10^{-7}$\\
\hline
Time interval $T^{\text{cloud}}$ & $[15,25]~\text{s}$ &The system parameter $\beta$ & 2\\
\hline
Computation capacity of each device $f_n$ & $[10^9, 4\times10^9]$ \text{cycles/s} & The finite accuracy loss $G$ & 3\\
\hline
\end{tabular}
\end{table}

We compare the proposed algorithm (Coalition altruistic preference rule) with three other benchmark schemes or rules. One is the Selfish preference rule, each device may frequently move
from the original coalition to a new coalition in the pursuit of high utility, which will hurt the interests of other devices. The second benchmark algorithm is Pareto preference rule, which can gradually improve the utility of the device while preserving the others utilities in the original and new coalitions.
The last benchmark algorithm is the Bandwidth optimization only algorithm, where where each device selects the coalition first according to its location, coalition reward and then the
bandwidth is allocated according to current coalition partition. The performance of the proposed algorithms
are evaluated by averaging over 800 experiments.

\begin{figure}[htbp]
\centering
\begin{minipage}[t]{0.5\linewidth}
\label{fig:An example of the final stable coalition in edge FL network}
\centering
\includegraphics[height=4.5cm,width=7.5cm]{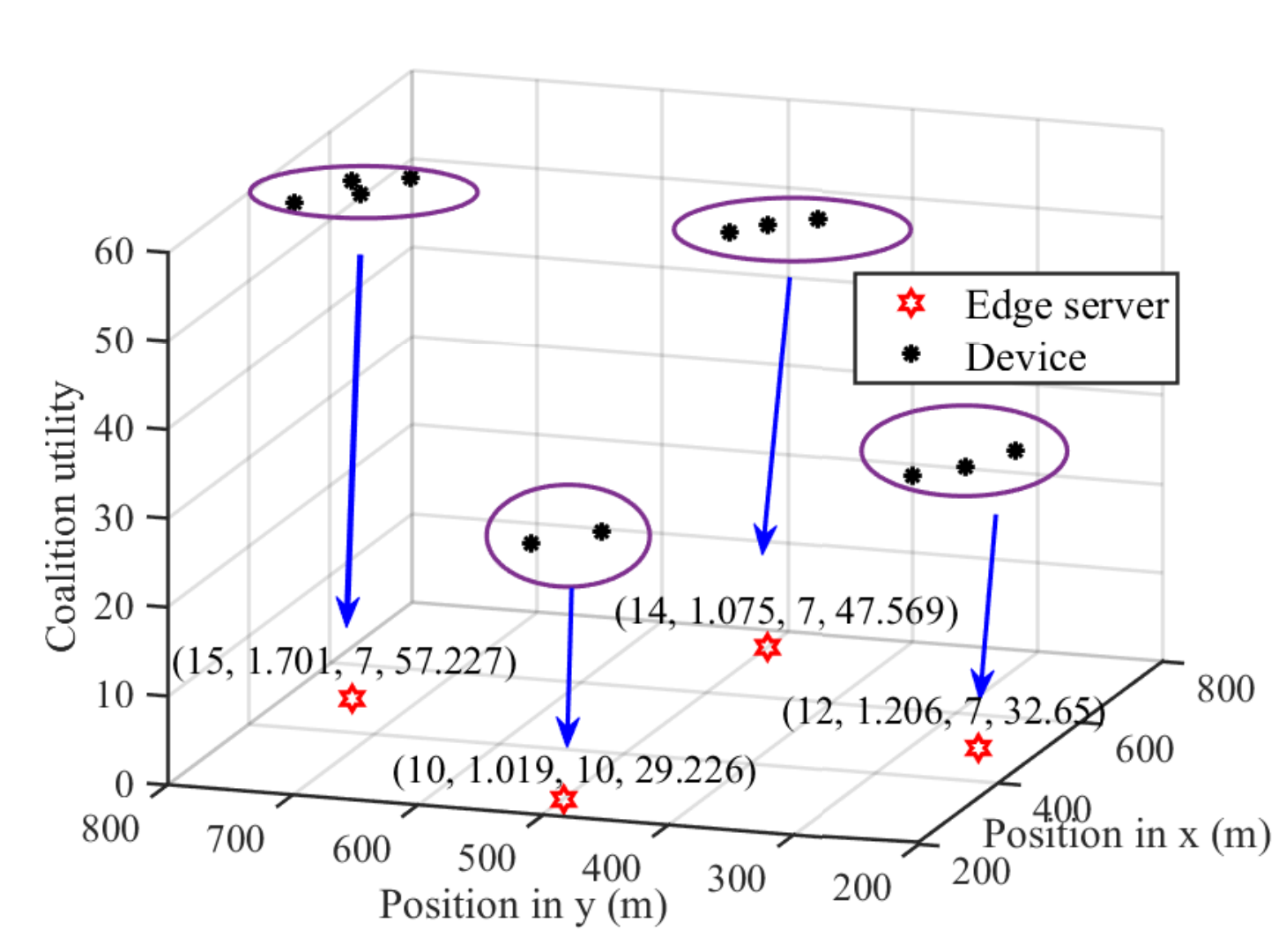}
\caption{An example of the final stable coalition in edge FL network.}
\end{minipage}%
\begin{minipage}[t]{0.5\linewidth}
\label{fig:The_converge_behavior_of_proposed_algorithm}
\centering
\includegraphics[height=4.5cm,width=7.5cm]{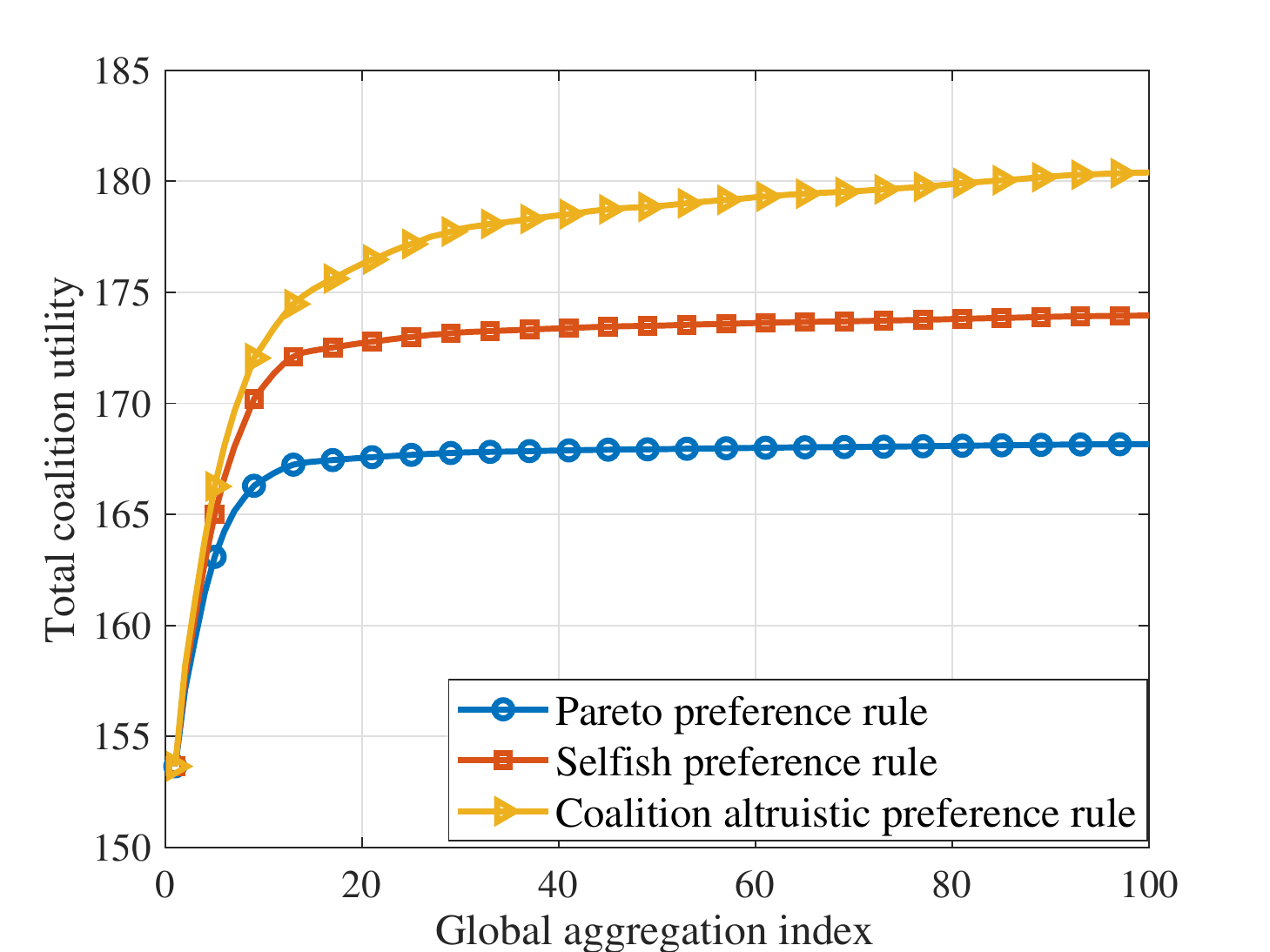}
\caption{The converge behavior of proposed algorithm based on three preference rules.}
\end{minipage}
\end{figure}


A demonstration of the final stable coalition in the edge FL network is presented in Fig~2, where $12$ devices within the HFL network are associated with edge servers to form $4$ stable coalitions. Each stable coalition has its own attribute. For example, for coalition 1 in Fig~2, its attribute is $(15,1.701,7,57.227)$, in which $\rho_1=15$ represents the unit price provided by the edge server for the model improvement brought on by the coalition's local training, $B_1= 1.701$ MHz means the optimal bandwidth allocated to the coalition according to Algorithm 2, $K_1=7$ is the number of edge aggregations executed by devices in the coalition, and $u^{\mathcal{S}_1}=57.227$ is the utility of all devices in the coalition. Due to the highest rewards provided by the edge server $1$, more devices are willing to provide local training for that server, resulting in the highest utility for coalition $1$. Moreover, to ensure communication efficiency for edge FL, the largest amount of bandwidth has been allocated to coalition $1$.


Fig.~3 depicts the convergence behavior, the total utilities of all devices in HFL network versus global aggregation index. To generate the following curves, we averaged the results of 800 separate trials. We noticed that the three preference orders can converge to a stable state, in which the devices in coalitions adopt three different preferences, i.e., coalition expected altruistic preference rule, selfish preference rule, and Pareto preference rule, while the cloud server and edge servers always adopt Stackelberg games. Fig.~3 further demonstrates that while the coalition expected altruistic order converges a little slower than the selfish and Pareto orders, it can yield the highest total coalition utility. By contrast, the selfish order only prioritizes its own utility, which could negatively impact other coalitions' utilities. The coalition expected altruistic order allows for partial cooperation among devices, resulting in the best overall utility despite a minor decrease in convergence rate.

\begin{figure}[htbp]
\centering
\begin{minipage}[t]{0.50\linewidth}
\label{fig:The_total_utility_of_all_coalitions_versus_the_time_interval}
\centering
\includegraphics[height=4.5cm,width=7.5cm]{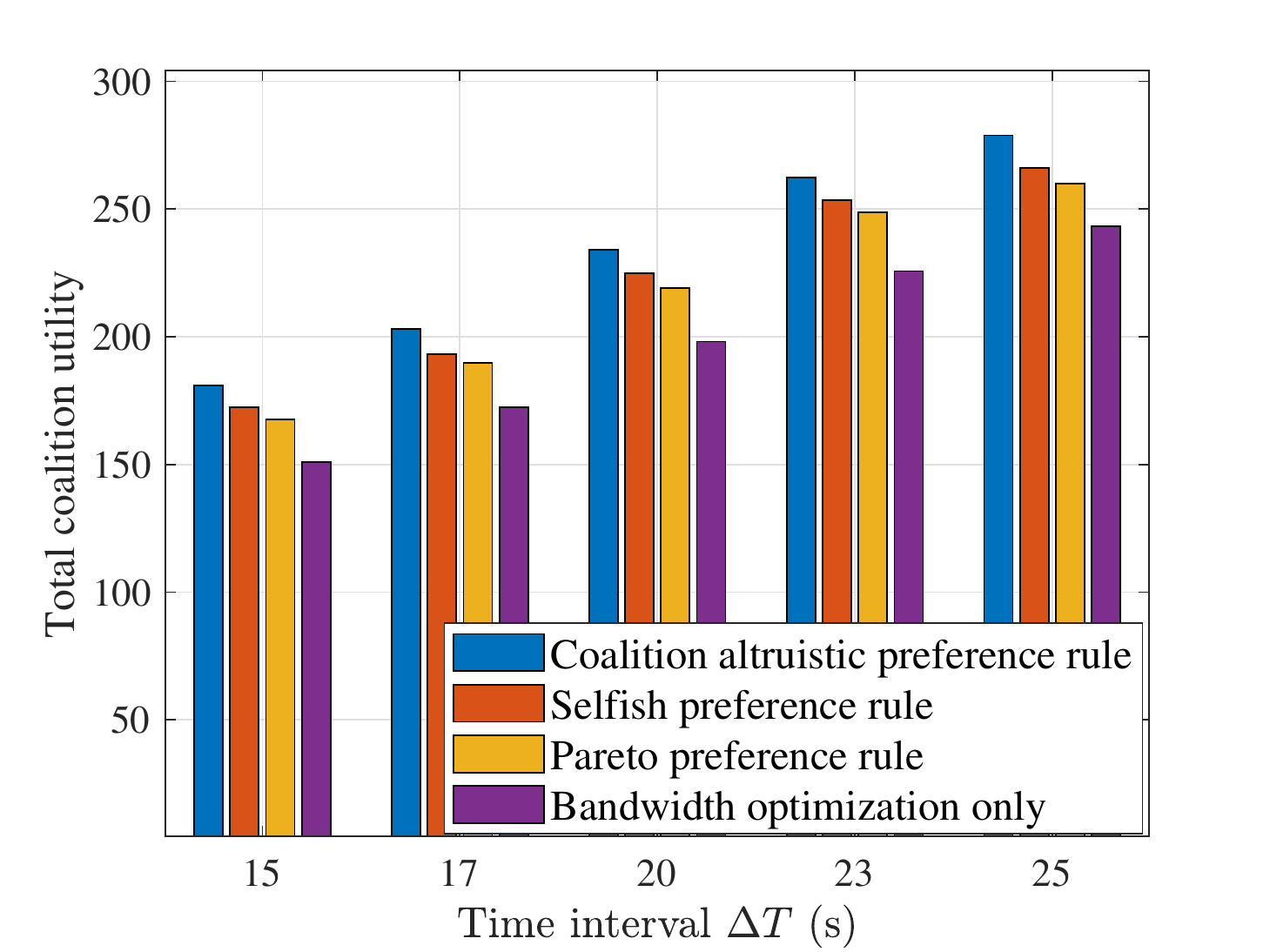}
\caption{The total coalition utilities versus the time interval of global aggregation.}
\end{minipage}%
\begin{minipage}[t]{0.50\linewidth}
\label{fig:The_total_utility_of_all_coalitions_versus_the_number_of_devices1}
\centering
\includegraphics[height=4.5cm,width=7.5cm]{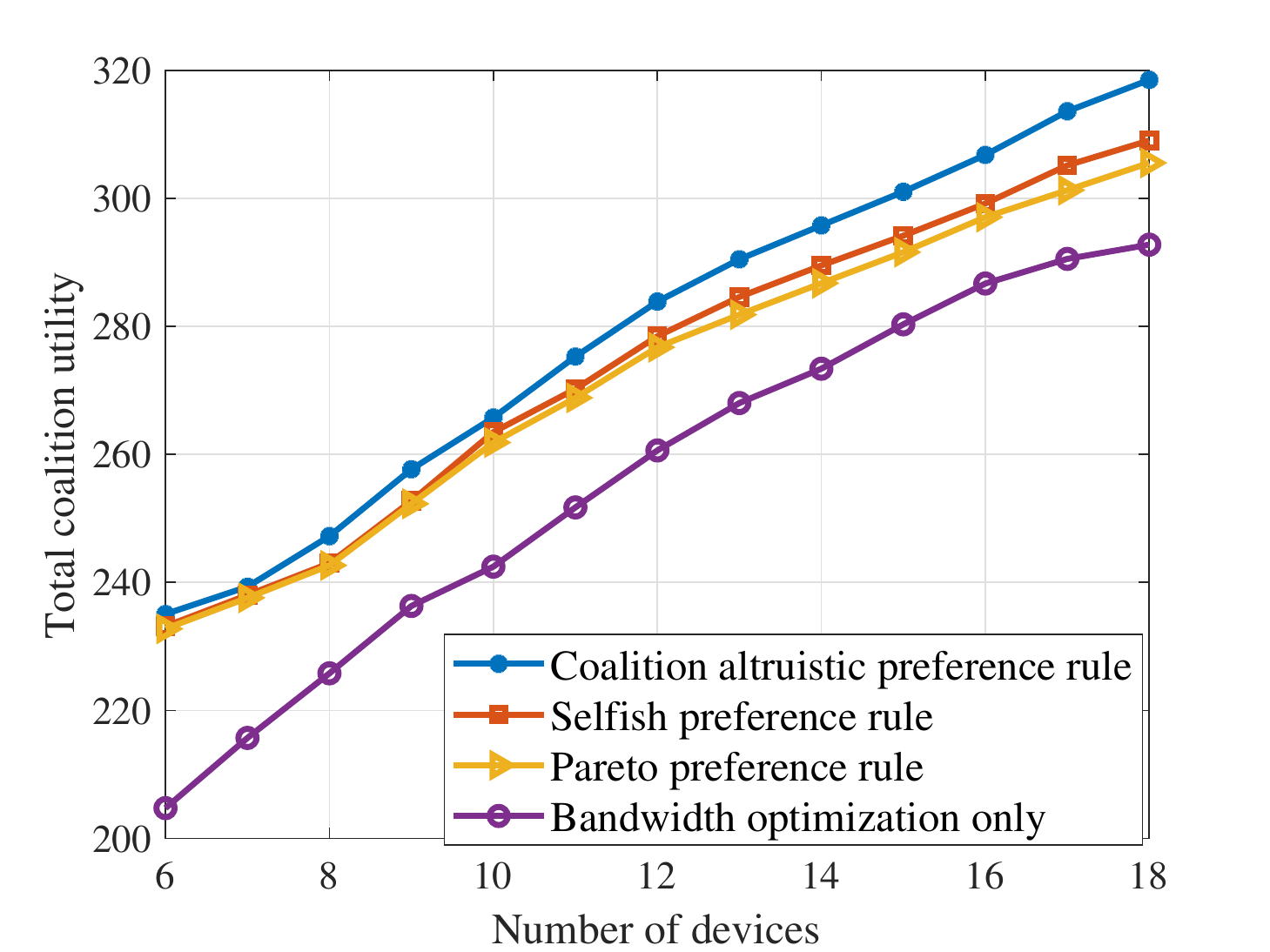}
\caption{The total coalition utilities versus the number of devices with low communication overhead.}
\end{minipage}
\end{figure}


In Fig.~4, we evaluate the performance of the different preference rules for a HFL network having
$4$ edge servers and $12$ devices, as the time interval of global aggregation increases. As shown in Fig.~4, it is evident that the overall utility of all devices are bound to increase as the time interval of global aggregation increase in HFL network. The reason for this is that longer time interval of global aggregation lead to more sufficient local training of the devices, resulting in higher utilities. Moreover, the utility achieved by the coalition expected altruistic preference rule is better than that of two traditional orders, and the Bandwidth optimization only algorithm exhibits the poorest performance among all. Note that the Pareto preference rule is close to
Bandwidth optimization only algorithm due to its the strong restriction.



In Fig.~5 and~6, we assess the total utilities of coalitions versus the number of devices. We compare the utilities between the four schemes under different communication overhead $\alpha_l$, in which the communication costs that each device needs to pay in Fig.~5 are lower, whereas the communication costs for each device in Fig.~6 are relatively higher. It is apparent that the achieved total utility under coalition expected altruistic preference rule is higher than that of other three schemes or rules regardless of the communication costs. In Fig.~5, the utility of all coalition rely more on local data for training when the communication cost is relatively low, and involving more devices in the local training leads to an increase in training data. Hence, the total utility of the coalitions increase as the number of devices participating in the local training increases with low communication costs. However, the curves in Fig.~6 first increase to the maximum points and then decrease as the number of devices increase. When utility increases, it represents the benefits of cooperation, whereas a decrease in utility indicates that the costs of cooperation outweigh the revenue gained.


\begin{figure}[htbp]
\centering
\begin{minipage}[t]{0.50\linewidth}
\label{fig:number_of_devices2}
\centering
\includegraphics[height=4.5cm,width=7.5cm]{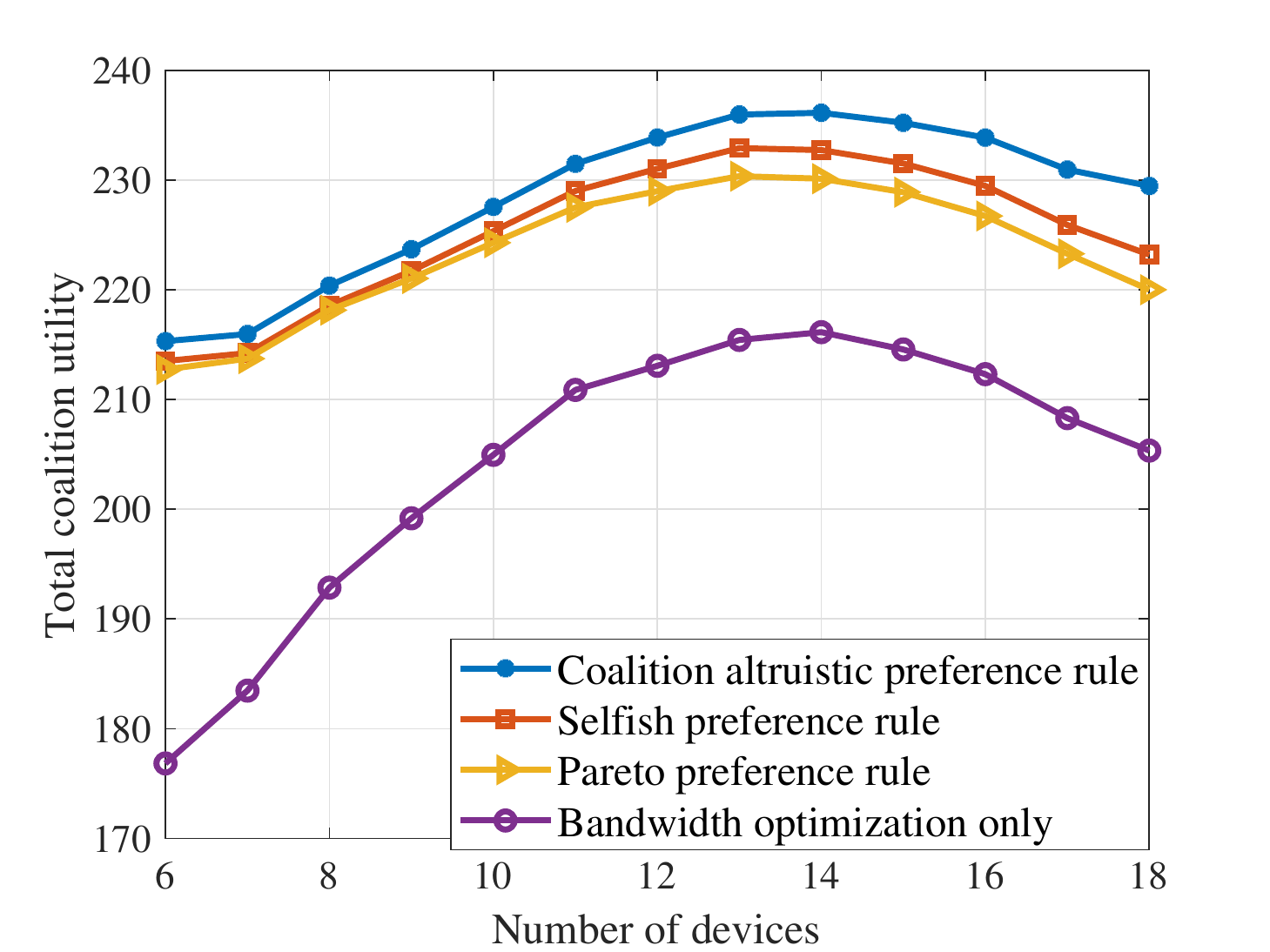}
\caption{The total coalition utilities versus the number of devices with high communication overhead.}
\end{minipage}%
\begin{minipage}[t]{0.50\linewidth}
\label{fig:utility_of_server_verse_server}
\centering
\includegraphics[height=4.5cm,width=7.5cm]{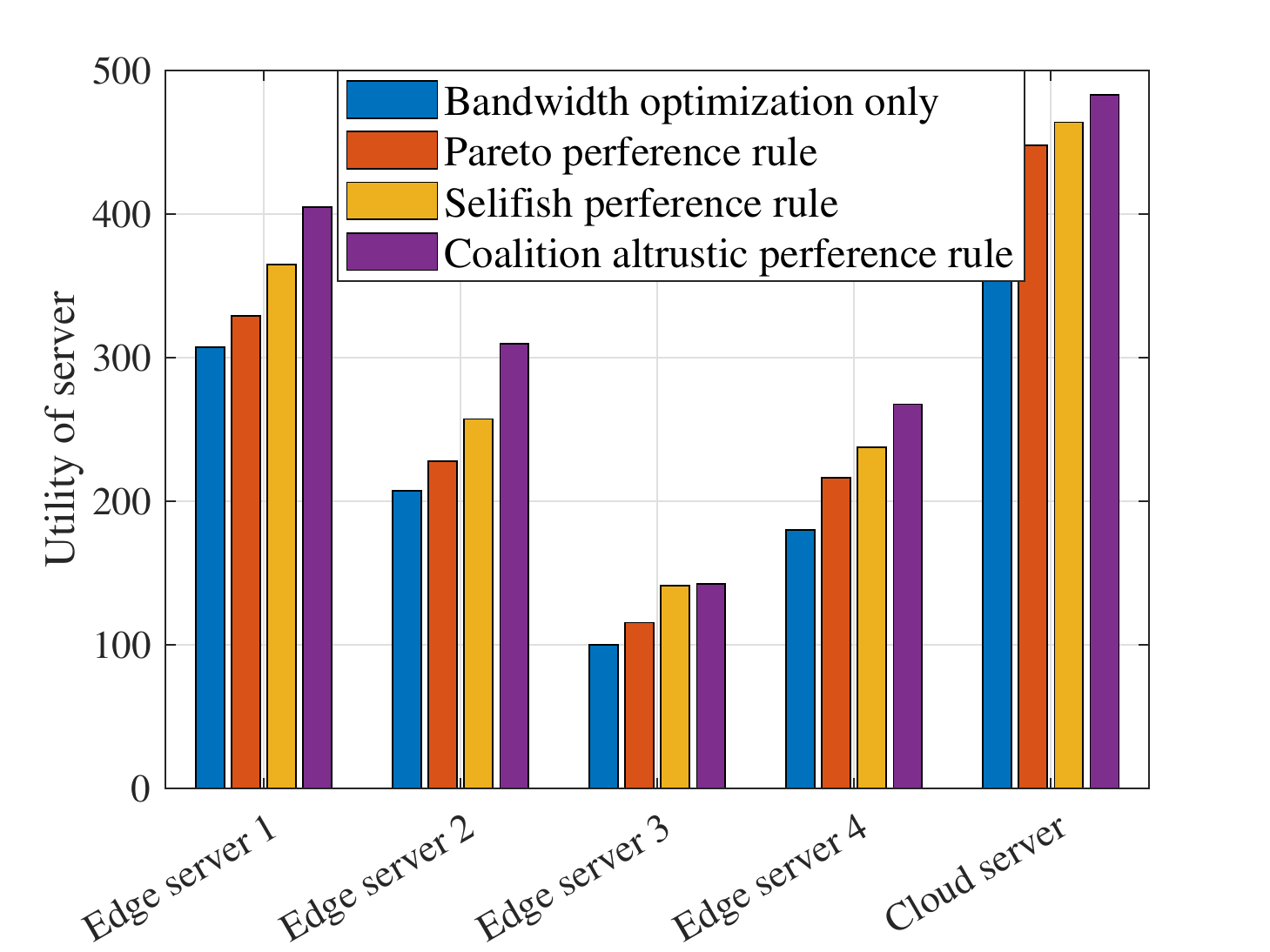}
\caption{The utilities of each server in HFL network}
\end{minipage}
\end{figure}

In Fig.~7, we evaluate the performance of edge servers and cloud server under different preference rules or schemes for a network having 4 edge servers and 12 devices. Each edge server (coalition) attribute is given in Fig.~7. It is obviously that the coalition altruistic preference rule can bring higher utility to those servers compared to the other three schemes. This is because the coalition altruistic preference rule forms more effective coalition partitions for all devices, thereby enhancing the utility of participants in the upper-level game.

\begin{figure}[htbp]
\centering
\begin{minipage}[t]{0.50\linewidth}
\centering
\includegraphics[height=4.5cm,width=7.5cm]{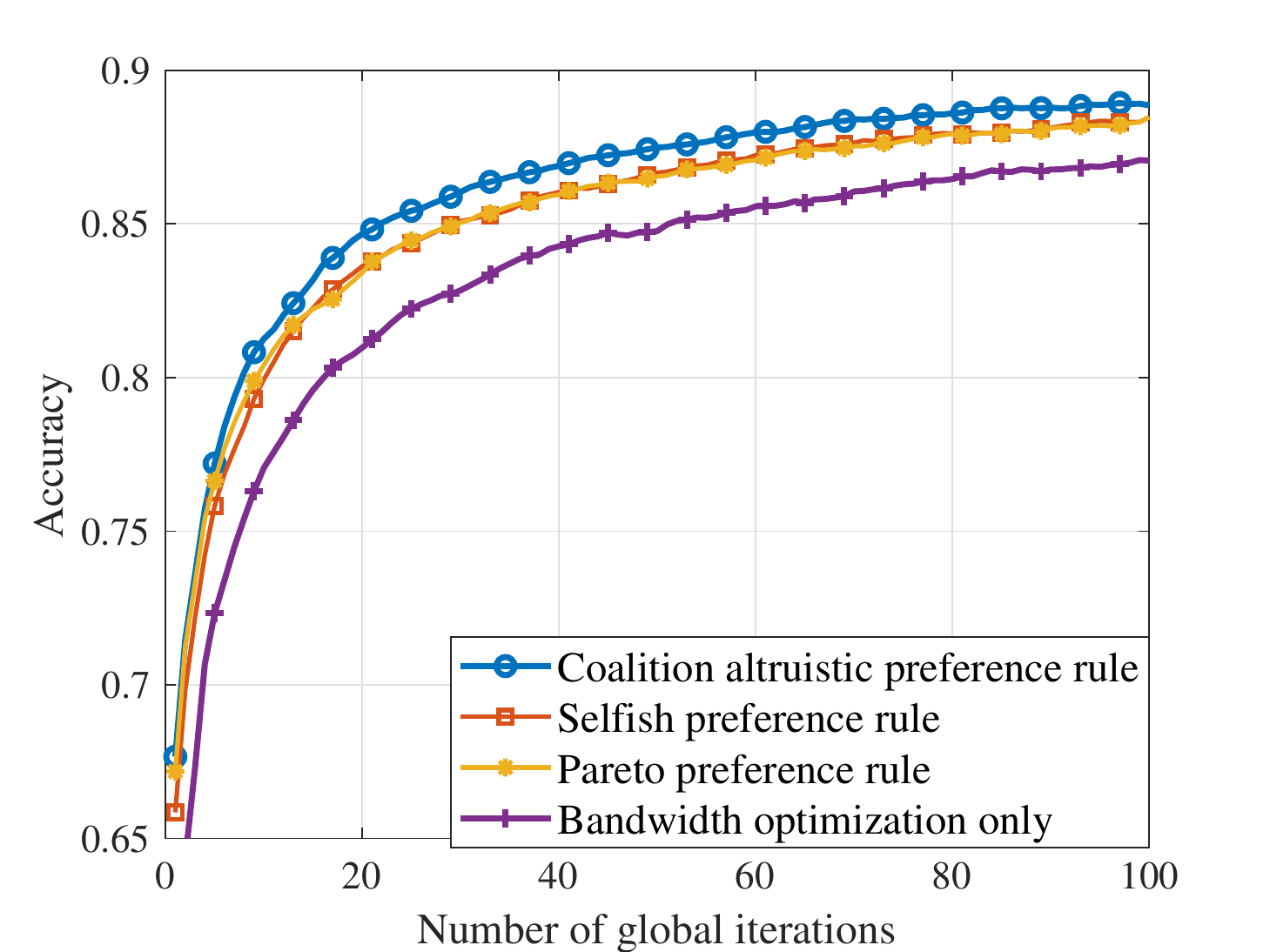}
\caption{The HFL accuracy versus the number of global aggregations on FashionMNIST data.}
\label{fig:Fmnist10_acc}
\end{minipage}%
\begin{minipage}[t]{0.50\linewidth}
\centering
\includegraphics[height=4.5cm,width=7.5cm]{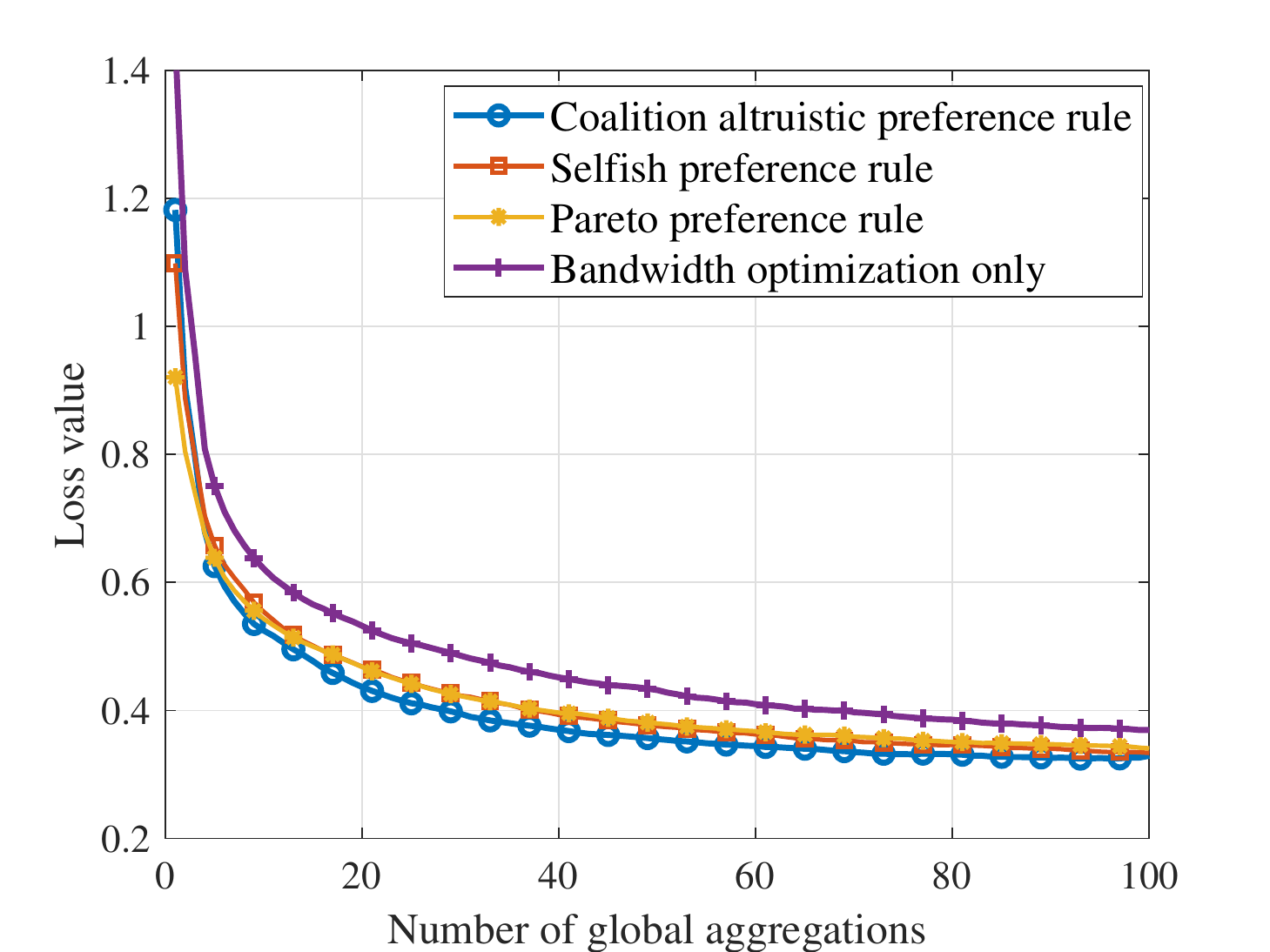}
\caption{The HFL loss value versus the number of global aggregations on FashionMNIST data.}
\label{fig:Fmnist10_loss}
\end{minipage}
\end{figure}

%
\begin{figure}[htbp]
\centering
\begin{minipage}[t]{0.50\linewidth}
\centering
\includegraphics[height=4.5cm,width=7.5cm]{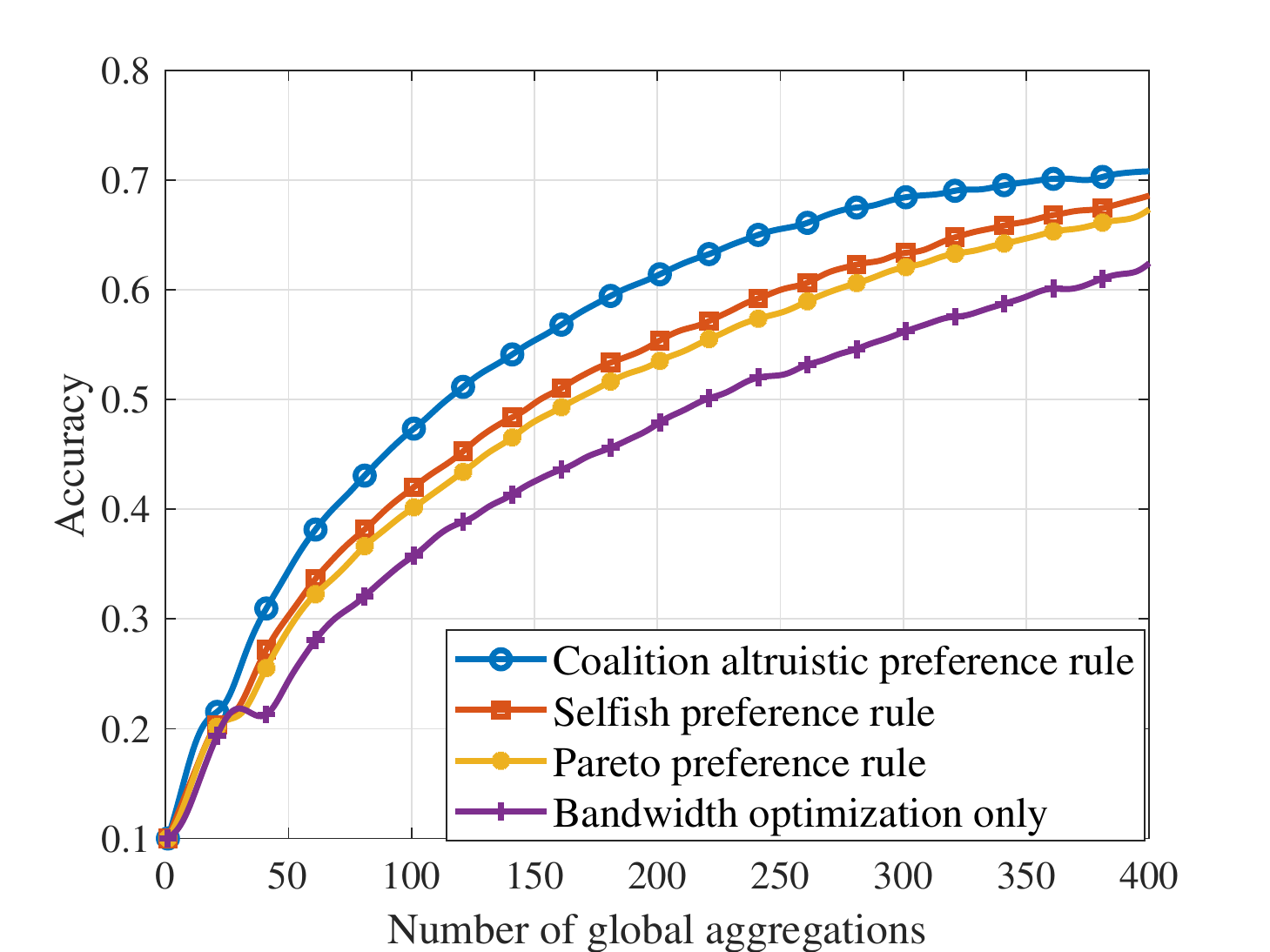}
\caption{The HFL accuracy versus the number of global aggregations on Cifar-10.}
\label{fig:cifar_acc}
\end{minipage}%
\begin{minipage}[t]{0.50\linewidth}
\centering
\includegraphics[height=4.5cm,width=7.5cm]{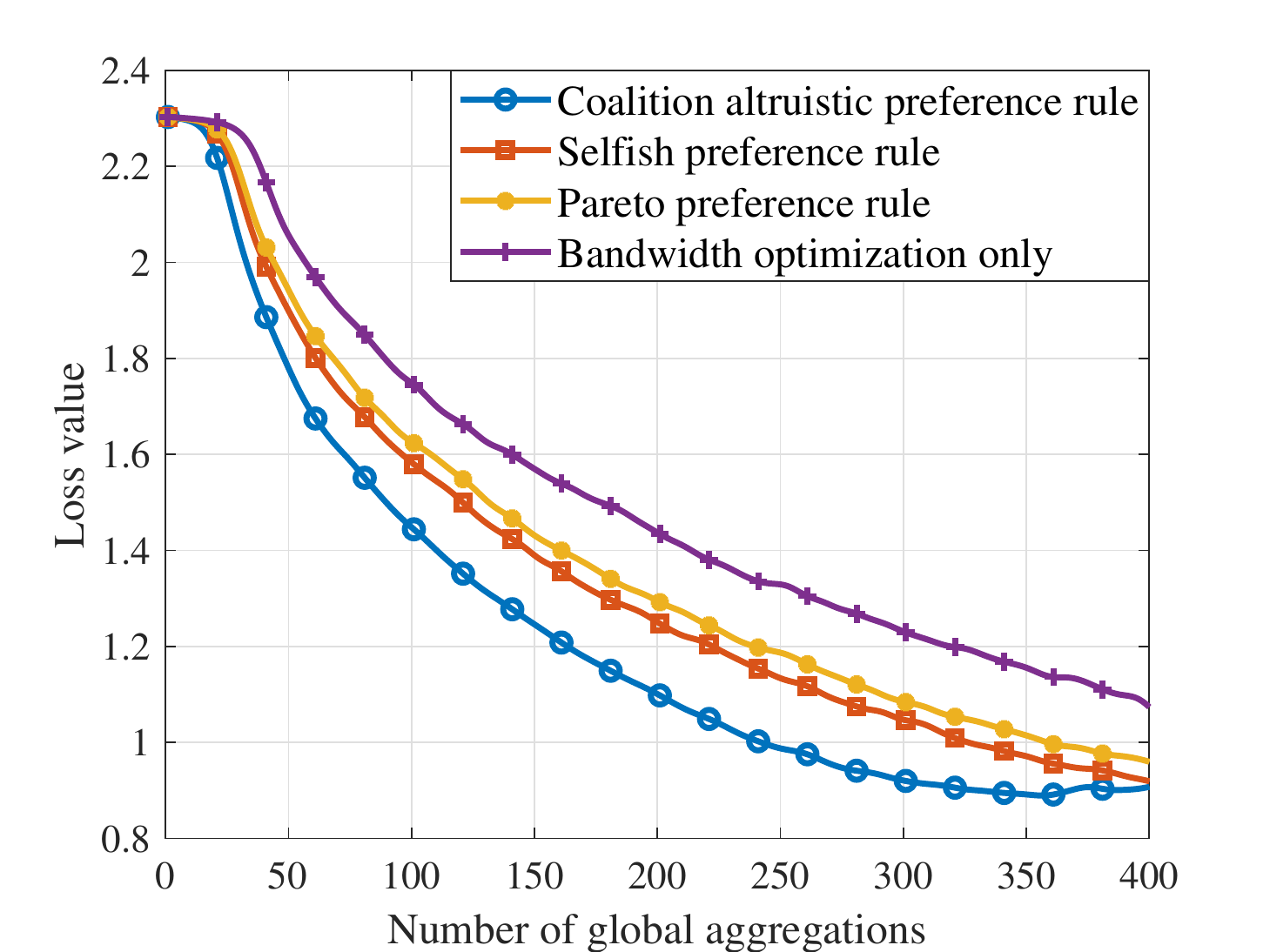}
\caption{The HFL loss value versus the number of global aggregations on Cifar-10.}
\label{fig:cifar_loss}
\end{minipage}
\end{figure}

%

In Fig.~\ref{fig:Fmnist10_acc} and~\ref{fig:Fmnist10_loss}, we show how the HFL accuracy and the loss value changes as the number of global aggregations varies on FashionMNIST dataset. From Figs.~\ref{fig:Fmnist10_acc} and~\ref{fig:Fmnist10_loss}, we can see that the coalition expected altruistic preference rule is significantly better than the other schemes, and its convergence speed is also significantly faster than three other benchmark schemes or rules. Due to the relatively small size of the FashionMNIST dataset and its ease of training, there is not a significant performance gap between selfish preference rule and Pareto preference rule. Figs.~\ref{fig:cifar_acc} and \ref{fig:cifar_loss} show how the HFL accuracy and the loss value changes with the number of global aggregations on Cifar-10 dataset.  We can also see that, the coalition expected altruistic preference rule can achieve up to 3\% gains in
terms of the accuracy compared with three other benchmark schemes or rules. As seen from Fig.~\ref{fig:Fmnist10_acc} to~\ref{fig:cifar_loss}, the Bandwidth optimization only algorithm has lowest learning performance and speed, as well as poor stability owing to huge fluctuations compared with other algorithms.

\section{Conclusion}
\label{sec:conclusion}
In this paper, we first designed two-level incentive mechanisms based on game theory for the HFL in a device-edge-cloud coordinating architecture, aiming at encouraging the participation of entities in each level. In the design of lower-level incentive mechanism, we proposed a coalition formation game to optimize the device-edge association and bandwidth allocation. To be specific, we first developed efficient coalition partitions based on preference rules for optimizing device-edge association, which can be proven to be stable by exact potential function. Then, we developed a gradient projection method to optimally allocate bandwidth among the coalitions. In the upper-level game, we designed a Stackelberg game algorithm to maximize the utilities of the cloud and each edge server. The proposed algorithm was able to determine the optimal number of aggregations at each edge server, as well as the reward provided by the cloud for the performance improvement due to aggregations of each edge server. Numerical results indicated that the proposed two-level incentive mechanisms can achieve better performance than the benchmark schemes. Additionally, the mechanism we proposed demonstrated superior performance on real datasets.


%
 \appendices
 \section{Proof of Theorem \ref{theorem 3}}
 \label{Appendix A}
According to \cite{zhang2018context}, we denote the potential function as,
\begin{equation}
\psi(a_n, a_{-n})=\sum_{l \in \mathcal{L}}\sum_{i \in \mathcal{S}_l} u_{i}^{\mathcal{S}_l}(a_n, a_{-n}),
\end{equation}
which is the sum of the utilities of all devices from all coalitions. When the device $n$ changes its strategy, the difference of the potential function is
\begin{equation}
\label{difference of the potential function}
\begin{split}
&\psi(a_n, a_{-n})-\psi(\widetilde{a_n}, a_{-n})
=\sum_{l \in \mathcal{L}}\sum_{i \in \mathcal{S}_l} u_{i}^{\mathcal{S}_l}(a_n, a_{-n})- \sum_{l \in \mathcal{L}}\sum_{i \in \mathcal{S}_l} u_{i}^{\mathcal{S}_l}(\widetilde{a_n}, a_{-n})\\
&=\sum_{i \in \mathcal{S}_{a_n}\setminus \{n\}} \left[u_{i}^{\mathcal{S}_{a_n}}(a_n, a_{-n}) -u_{i}^{\mathcal{S}_{a_n}}(\widetilde{a}_n, a_{-n})\right]
+\sum_{i \in \mathcal{S}_{\widetilde{a}_n}\setminus \{n\}} \left[u_{i}^{\mathcal{S}_{\widetilde{a}_n}}(a_n, a_{-n}) -u_{i}^{\mathcal{S}_{\widetilde{a}_n}}(\widetilde{a}_n, a_{-n})\right]\\
&+\sum_{i \in \mathcal{S}_{o.w.}} \left[u_{i}^{\mathcal{S}_{o.w.}}(a_n, a_{-n}) -u_{i}^{\mathcal{S}_{o.w.}}(\widetilde{a}_n, a_{-n})\right]
+u_{n}^{\mathcal{S}_{a_n}}(a_n, a_{-n})-u_{n}^{\mathcal{S}_{\widetilde{a}_n}}(\widetilde{a}_n, a_{-n}),
\end{split}
\end{equation}
where $\mathcal{S}_{o.w.}$ means $\{\mathcal{N}\setminus \{\mathcal{S}_{a_n} \bigcup \mathcal{S}_{\widetilde{a}_n}\}\}$. According to the \textit{Definition} \ref{definition 7} and \ref{definition 8}, we know that other coalitions $\mathcal{S}_{o.w.}$ will be unaffected by device $n$'s strategy change, and the forth part is $0$.
Recall (\ref{coalition altruistic preference rule}) in \textit{Definition} \ref{definition 6}, the utility of device $n$ can be expressed as $U_n(a_n, a_{-n})=\sum_{i \in \mathcal{S}_{a_n}\setminus \{n\}} u_{i}^{\mathcal{S}_{a_n}}(a_n, a_{-n})+\sum_{i \in \mathcal{S}_{\widetilde{a}_n}\setminus \{n\}} u_{i}^{\mathcal{S}_{\widetilde{a}_n}}(a_n, a_{-n})+u_{n}^{\mathcal{S}_{a_n}}(a_n, a_{-n})$The difference in utility function is described as follow:
\begin{equation}
\label{difference in utility function}
\begin{split}
&U_n(a_n, a_{-n})- U_n(\widetilde{a}_n, a_{-n})=\sum_{i \in \mathcal{S}_{a_n}\setminus \{n\}} \left[u_{i}^{\mathcal{S}_{a_n}}(a_n, a_{-n}) -u_{i}^{\mathcal{S}_{a_n}}(\widetilde{a}_n, a_{-n})\right]\\
&+\sum_{i \in \mathcal{S}_{\widetilde{a}_n}\setminus \{n\}} \left[u_{i}^{\mathcal{S}_{\widetilde{a}_n}}(a_n, a_{-n}) -u_{i}^{\mathcal{S}_{\widetilde{a}_n}}(\widetilde{a}_n, a_{-n})\right] +u_{n}^{\mathcal{S}_{a_n}}(a_n, a_{-n})-u_{n}^{\mathcal{S}_{\widetilde{a}_n}}(\widetilde{a}_n, a_{-n})\\
&=\psi(a_n, a_{-n})-\psi(\widetilde{a_n}, a_{-n}).
\end{split}
\end{equation}
Hence, the coalition game as an exact potential game has at least one pure strategy Nash equilibrium. \textit{Theorem} \ref{theorem 3} has been proved.
$\hfill\blacksquare$

%

\ifCLASSOPTIONcaptionsoff
\newpage
\fi


\bibliographystyle{IEEEtran}
\bibliography{ANN}





\end{document}